\documentclass[a4paper,11pt]{article}
\usepackage{jheppub} 
\usepackage{lineno}
\usepackage{macros}
\IfFileExists{stmaryrd.sty}{\usepackage{stmaryrd}}{}
\usepackage{theorems}
\usepackage{mathtools}
\usepackage{booktabs}
\usepackage{cleveref}

\graphicspath{{figures/}}

\definecolor{CBBlack}{RGB}{187,187,187}
\definecolor{CBOrange}{RGB}{230,159,0}
\definecolor{CBSkyBlue}{RGB}{86,180,233}
\definecolor{CBGreen}{RGB}{0,158,115}
\definecolor{CBYellow}{RGB}{240,228,66}
\definecolor{CBBlue}{RGB}{0,144,178}
\definecolor{CBBrown}{RGB}{213,94,0}
\definecolor{CBPurple}{RGB}{204,121,167}

\newcommand{\Tr}[0]{\mathrm{Tr}}
\newcommand{\dd}[0]{\mathrm{d}}
\newcommand{\pDM}[0]{\rho_{\m{DM}}}
\newcommand{\pDE}[0]{\rho_{\m{DE}}}
\newcommand{\wDM}[0]{\Omega_{\m{DM}}}
\newcommand{\wDE}[0]{\Omega_{\m{DE}}}
\newcommand{\swDM}[0]{\overline{\Omega}_{\m{DM}}}
\newcommand{\swDE}[0]{\overline{\Omega}_{\m{DE}}}
\newcommand{\wB}[0]{\Omega_{\m{b}}}
\newcommand{\wR}[0]{\Omega_{\m{r}}}
\newcommand{\wV}[0]{\Omega_{V}}
\newcommand{\weff}[0]{w_{\m{eff}}}
\newcommand{\sweff}[0]{\overline{w}_{\m{eff}}}
\newcommand{\lb}[0]{\lambda_{\m{b}}}
\newcommand{\lr}[0]{\lambda_{\m{r}}}

\title{Cosmological stasis and the coupled dark sector of the Dark Dimension}

\author[a]{Alexander Stewart}
\emailAdd{ajstewa1@uci.edu}

\affiliation[a]{Department of Physics and Astronomy, University of California, Irvine, CA}

\abstract{We study cosmological stasis in coupled dark matter and dark energy models in which a scalar field drives the dark energy and sets the dark matter mass, with the Dark Dimension as the motivating realization. Treating dark matter, dark energy, baryons, and radiation as an autonomous dynamical system, we show that exact stasis requires exponential slopes along the trajectory, and that running slopes give exact stasis or a drifting quasi-stasis depending on whether the running is integrable. The constant abundance solutions are the stationary solutions of coupled quintessence, which we read as stasis epochs with computable lifetimes and exits. Anchoring the cosmology to the measured abundances at matter-radiation equality and today, we show that the stasis region for this model cannot accelerate for any initial conditions, and that the total equation of state is never phantom although the effective dark energy of a $\Lambda$CDM fit can appear so. For the Dark Dimension, capture into stasis in the future is excluded for geometric sources of the potential once dark fifth-force bounds are imposed, and dark matter production from the Standard Model brane supports a further stasis which the standard scenario does not reach. The stasis solutions provide analytic exclusions of the parameter space which do not depend on initial conditions.}

\begin{document}
\maketitle
\flushbottom

\section{Introduction}\label{sec:intro}

Recent DESI DR2 results \cite{DESI:2025fii,DESI:2025zgx} have challenged the $\Lambda$CDM standard framework and suggest an evolving dark energy sector, see \cite{Ramadan:2024kmn,Gialamas:2025pwv,Akrami:2025zlb,Anchordoqui:2025fgz,Bayat:2025xfr,Berbig:2024aee,deSouza:2025rhv,Borys:2026sna} for quintessence fits to these data. Coupling between dark matter and dark energy sectors allows for an evolving dark sector which can be in agreement with new cosmological data. Fluid level interactions with an energy transfer proportional to the Hubble rate have been fit to the DESI data in \cite{Giare:2024smz,Li:2024qso,Silva:2025hxw,Pan:2025qwy,Li:2025ula,vanderWesthuizen:2025iam,vanderWesthuizen:2025rip,Petri:2025swg,Figueruelo:2026eis}, and scalar field models in which the coupling follows from a field dependent dark matter mass or from a momentum exchange in \cite{Pourtsidou:2025sdd,Bedroya2025,Chen:2025ywv,Aoki:2025bmj,Antusch:2026ldp,Samanta:2025oqz,Khoury:2025txd,Chen:2025wwn,GomezValent:2026dea,Zhai:2026uwr}. In particular, \cite{Bedroya2025} shows that a coupled dark matter and dark energy arising naturally from a dynamical extra dimension in the Dark Dimension (DD) scenario \cite{Montero:2022prj} can fit the current cosmological data on an evolving dark energy sector.

Motivated by this model, we study whether such a coupled dark sector can exhibit cosmological stasis and use the resulting solutions to bound the parameter space of such a model. Cosmological stasis is an extended period of time in which some or all cosmological abundances remain constant even though the universe is expanding \cite{Dienes:2021woi}. Components with different equations of state redshift at different rates, so constant abundances require a process which continually transfers energy from the components which would otherwise grow to the components which would otherwise fade. In the original proposal \cite{Dienes:2021woi} this process occurs through the decay of a tower of massive states into radiation, and stasis arises by the balance of the decay of matter into radiation against the faster dilution of radiation as compared to matter. The same balance has since been found with other sources of energy transfer, including the evaporation of a spectrum of primordial black holes \cite{Dienes:2022zgd}, the dynamics of scalars with time dependent equations of state \cite{Dienes:2024wnu}, thermal annihilations of a single species \cite{Barber:2024vui}, a field dependent decay width \cite{Huang:2025odd}, and gravitational production of towers in the early universe \cite{Long:2025wjw}, and the interplay of annihilation and decay \cite{Barber:2024izt}. The structure common to all of these was analyzed for two and three components in \cite{Dienes:2023ubz}, and the generality of the phenomenon was explored with machine learning techniques in \cite{Halverson:2024oir}. Two features of this body of work are relevant here. First, the transfer process always involves particle physics, in the form of decays, annihilations, or production. Second, all of these stasis epochs occur in the early universe, before big-bang nucleosynthesis (BBN), and end when the tower is exhausted.

The coupled dark sector supplies a transfer process of a different kind. The energy flows between dark matter and dark energy through the field dependence of the dark matter mass, not through any decay, annihilation, or production process. Furthermore, there is no reason for these stasis solutions to be confined to the early universe. The question is then whether this exchange can balance the expansion in the same way, and if so, whether the resulting stasis could be relevant to various periods in the evolution of our universe.

Solutions of coupled quintessence with constant abundances have in fact been known for a long time. The coupled scaling solution was found in \cite{Wetterich:1994bg}, the critical points of the coupled system were classified in \cite{Amendola:1999er}, see also \cite{Pettorino:2008ez}, and the classification for the same four component cosmology studied here, including uncoupled baryons and radiation, was carried out in \cite{Amendola:2000ub,Amendola:2001rc} (which builds upon the uncoupled case originally studied in \cite{Copeland:1997et,Ferreira:1997hj}). See \cite{Copeland:2006wr,Bahamonde:2017ize} for reviews. To our knowledge these stationary solutions have not been examined as realizations of cosmological stasis. Viewing them this way raises questions which the scaling literature did not ask. One would like to know which potentials admit such solutions, how long the solutions persist, what ends them, and whether our universe has been in, is in, or will be in one of them.

Related questions for uncoupled multi-exponential potentials, in particular which scaling attractors accelerate independently of initial conditions, were studied in \cite{Shiu:2023fhb}. In this paper we answer these questions for general $V(\phi)$ and $M(\phi)$ and then specialize to the DD realization of the model. There $\phi$ is the radion of a mesoscopic extra dimension, the dark matter is a tower of Kaluza-Klein (KK) gravitons \cite{Gonzalo:2022jac,Anchordoqui:2022svl} (see \cite{Dienes:2011ja} for related work on dynamical DM, and DM realized via KK towers), and both the potential and the tower mass scale depend on $\phi$ \cite{Agrawal:2019dlm,Bedroya2025}. \Cref{sec:model} writes the cosmology of dark matter, dark energy, baryons, and radiation as an autonomous dynamical system. \Cref{sec:theorem} shows that exact stasis forces the field onto a straight trajectory along which $M$ and $V$ must be exponential wherever dark matter and potential energy are present, which extends \cite{Liddle:1998xm,Amendola:2006qi} to the cosmology with spectators, and determines what happens when the slopes run. \Cref{sec:classification} classifies the constant abundance solutions of the exponential model and their stability, and \cref{sec:lifetimes} computes the lifetime and exit of the coupled stasis. \Cref{sec:obs} compares the solutions to observational data. We impose the dark fifth-force bound and show that the radiation era solutions have no exit into the standard cosmology. After anchoring the cosmology to the measured abundances at matter-radiation equality and today, we derive an exact bound on the present total equation of state, which is never phantom, together with the effective dark energy equation of state seen by a $\Lambda$CDM fit, which can appear phantom. The stasis region of this model cannot reproduce the observed acceleration, and geometric sources of the potential cannot capture the radion into stasis in the future once the fifth-force bound is imposed. \Cref{sec:production} includes the production of dark matter from the cooling of the Standard Model (SM) brane and finds a new stasis sustained by the drift of the tower mass, which the DD scenario as formulated does not reach. \Cref{sec:swamp} points out that the structure of the solution space lines up with several swampland bounds, and \cref{sec:conc} concludes along with a discussion on how our model can be generalized to re-open regions of the stasis parameter space. Throughout, we work in reduced Planck units such that $M_{\m{pl}}^2 = (8\pi G)^{-1} = 1$, and dots will denote derivatives with respect to time.

\section{The model}\label{sec:model}

\subsection{Setup}\label{sec:setup}
Consider the case where the scalar $\phi$ with potential $V(\phi)$ responsible for dark energy also affects the mass scale of dark matter, $M(\phi)$. Such a coupling appears naturally in extra dimensional theories where the moduli for the size of extra dimensions have non-trivial potentials and set the mass scale for KK excitations, see \cite{Basile:2025lek} for a braneworld realization of dynamical dark energy in string theory. We will keep in mind a specific example of the dark dimension scenario where the SM is taken to live on a $3$-brane, and there is one large extra dimension which leads to a light tower of KK modes which can act as dark matter. In \cite{Gonzalo:2022jac}, it was shown that in the DD scenario if the tower of KK modes were excitations of the graviton, then the cooling of the SM $3$-brane can populate this tower and provide phenomenologically viable explanations for the dark matter abundance today. In addition, the universal coupling of the graviton to the SM brane allows for interactions with the SM. It was expected that intra-tower decays of the KK gravitons to lower KK levels would outpace decays to the SM provided the mesoscopic extra dimension is sufficiently smooth \cite{Gonzalo:2022jac}. This would ensure that over cosmological epochs there is not substantial decay of dark matter to the SM sector. However, in \cite{Langhoff:2026wiz,Lee:2026myr} it has been shown that the assumed intra-tower decay rate is significantly suppressed relative to the previously assumed rates, which excludes a flat DD. The inclusion of warping is claimed to re-open some regions of parameter space for the DD with KK gravitons as DM, as mentioned in \cite{Lee:2026myr}. In this work however, we will focus on the case where the DM is indeed comprised of a tower of KK modes whose decays into the SM are suppressed with the caveat that such a model cannot be realized by a flat DD with KK gravitons. Instead, the dark matter and dark energy sectors exchange energy solely through the evolution of the scalar field $\phi$, in this case $\phi$ being the radion for the extra dimension. We then assume that the dark matter energy density is affected solely by the expansion of the universe and evolution of $\phi$.

Consider a flat background FLRW cosmology. In addition to the coupled dark sector, we include baryonic matter and radiation. The SM lives on the brane and does not couple to the radion at leading order, so the baryons are uncoupled dust. The total energy density of the universe is taken to be
\begin{equation}
    \rho_{\m{tot}} = \pDE + \pDM + \rho_{\m{b}} + \rho_{\m{r}}\;,
\end{equation}
with
\begin{subequations}
\begin{equation}
    \pDM = \pDM^0 M(\phi) a^{-3}\;,
\end{equation}
\begin{equation}
    \pDE = \frac{1}{2}\dot{\phi}^2 + V(\phi)\;,
\end{equation}
\begin{equation}
    \rho_{\m{b}} = \rho_{\m{b}}^0 a^{-3}\;,
\end{equation}
\begin{equation}
    \rho_{\m{r}} = \rho_{\m{r}}^0 a^{-4}\;.
\end{equation}
\end{subequations}
$M(\phi)$ should be considered as the difference in mass between the tower states. Every state in the tower scales with $M(\phi)$, so $\pDM^0$ can be thought of as the sum over the initial mass of all states in the tower divided by $M(\phi_i)$, with $\phi_i$ the initial value of the scalar field. Since the dark matter does not decay to SM particles, the change in $\pDM$ comes solely from the dilution due to expansion and from the change in $M(\phi)$,
\begin{equation}\label{eq:DMfluid}
    \dot{\rho}_{\m{DM}} = -3H\pDM + \frac{\d \log M}{\d\phi}\,\dot{\phi}\,\pDM\;.
\end{equation}
The equation of motion for $\phi$ is
\begin{equation}\label{eq:phi_eom}
    \ddot{\phi} + 3 H \dot{\phi} + \frac{\d V}{\d\phi} + \pDM\,\frac{\d \log M}{\d\phi} = 0\;,
\end{equation}
and the total pressure is
\begin{equation}
    p_{\m{tot}} = \frac{1}{2}\dot{\phi}^2 - V(\phi) + \frac{1}{3}\rho_{\m{r}}\;.
\end{equation}
We define the abundances $\Omega_i = \rho_i / 3H^2$ and the effective equation of state
\begin{equation}
    \weff \equiv \frac{p_{\m{tot}}}{\rho_{\m{tot}}}\;.
\end{equation}
Differentiating $\wDM = \pDM/3H^2$ gives
\begin{equation}
    \dot{\Omega}_{\m{DM}} = \wDM\bigg[\frac{\dot{\rho}_{\m{DM}}}{\pDM} - 2\frac{\dot{H}}{H}\bigg]\;.
\end{equation}
The first term is simplified via \cref{eq:DMfluid}. For the second, the Friedmann equations of \cref{app:FEqns} combine in a flat universe to
\begin{equation}
    \dot{H} = -\frac{1}{2}\big(\rho_{\m{tot}} + p_{\m{tot}}\big) = -\frac{3}{2}H^2\big(1 + \weff\big)\;.
\end{equation}
The dilution term $-3H$ in \cref{eq:DMfluid} cancels against the corresponding term in $-2\dot{H}/H = 3H(1+\weff)$, and the dark matter abundance evolves as
\begin{equation}\label{eq:wDM_eom}
    \dot{\Omega}_{\m{DM}} = \wDM\bigg[\dot{\phi}\,\frac{\d\log M}{\d\phi} + 3\,\weff H\bigg]\;.
\end{equation}
The same cancellation occurs for every fluid, so each abundance evolves at the rate at which its energy density redshifts relative to the total energy density. This equation displays the two effects which compete in this model. The first term is the drift of the dark matter mass, while the second term measures how far the expansion of the universe is from that of a dust dominated universe.

It will be convenient to introduce the logarithmic slopes
\begin{equation}\label{eq:slopes}
    c_1(\phi) \equiv -\frac{\d\log V}{\d\phi}\;,\qquad
    c_2(\phi) \equiv -\frac{\d\log M}{\d\phi}\;.
\end{equation}
At this stage $c_1$ and $c_2$ are arbitrary functions of $\phi$. Note that a power series ansatz for $V$ and $M$ is equivalent to a power series for $c_1$ and $c_2$, which will be relevant when we promote $c_1$ and $c_2$ to arbitrary functions of $\phi$. We fix the orientation of $\phi$ such that the tower mass decreases toward positive $\phi$. The fit of the DD model to current data in \cite{Bedroya2025} prefers a coupling $c_2 \simeq 0.05 \pm 0.01$, which also sits inside the dark fifth-force window derived in \cref{sec:combined}. We therefore use $c_2 = 0.05$ as the benchmark coupling throughout, in the figures and in the anchored cosmologies of \cref{sec:obs}, and we will indicate where results depend on this choice. The dynamics are invariant under
\begin{equation}
    \phi \rightarrow -\phi\;,\qquad (\dot\phi,\, c_1,\, c_2) \rightarrow (-\dot\phi,\, -c_1,\, -c_2)\;,
\end{equation}
so only the relative sign of $c_1$ and $c_2$ is physical. In particular the case with $c_2 < 0$ and $c_1 > 0$ (i.e. the case relevant for a tower of winding modes in string theories) is equivalent to the branch with $c_1 < 0$ and $c_2 > 0$, which also follows from T-duality of string theories on $\mathbb{S}^1$.

\subsection{The autonomous system}\label{sec:autonomous}
To analyze the space of solutions systematically, we rewrite the cosmology as an autonomous dynamical system, following \cite{Copeland:1997et,Amendola:1999er,Amendola:2001rc}. Let $N = \log a$ be the number of e-folds. Primes denote derivatives with respect to $N$, and we define
\begin{equation}
    u \equiv \frac{\d\phi}{\d N} = \frac{\dot{\phi}}{H}\;.
\end{equation}
The kinetic and potential fractions of the dark energy are then $u^2/6$ and
\begin{equation}
    \wV \equiv \frac{V}{3H^2}\;,
\end{equation}
respectively. The Friedmann equation becomes the constraint
\begin{equation}\label{eq:constraint}
    \frac{u^2}{6} + \wV + \wDM + \wB + \wR = 1\;,
\end{equation}
which we use to eliminate $\wV$. Note that the dark energy abundance is
\begin{equation}\label{eq:DE_def}
    \wDE = \frac{u^2}{6} + \wV\;.
\end{equation}
We work with the kinetic and potential parts separately rather than with $\wDE$. The reason is that the dynamics do not close in $\wDE$ alone, since configurations with the same $\wDE$ but a different split between kinetic and potential energy have different pressures and do not evolve in the same manner. In addition, $u$ retains the sign of the rolling, which distinguishes the two branches of solutions we will discuss later. The effective equation of state is a function of the state variables,
\begin{equation}\label{eq:weff_state}
    \weff = \frac{u^2}{6} - \wV + \frac{\wR}{3} = \frac{u^2}{3} + \wDM + \wB + \frac{4}{3}\wR - 1\;.
\end{equation}
Converting \cref{eq:phi_eom} and the fluid equations to e-fold time (see \cref{app:aut_ODEs}) gives four coupled first order autonomous ODEs,
\begin{subequations}\label{eq:aut_sys_full}
\begin{align}
    u' &= -\frac{3}{2}\big(1 - \weff\big)u + 3\,c_1(\phi)\,\wV + 3\,c_2(\phi)\,\wDM\;,\label{eq:aut_u}\\
    \wDM' &= \big(3\,\weff - c_2(\phi)\,u\big)\wDM\;,\label{eq:aut_dm}\\
    \wB' &= 3\,\weff\,\wB\;,\label{eq:aut_b}\\
    \wR' &= \big(3\,\weff - 1\big)\wR\;.\label{eq:aut_r}
\end{align}
\end{subequations}
When the slopes depend on $\phi$ one must also include the equation $\phi' = u$. For constant slopes the equations for $u$ and the abundances close by themselves. Each abundance evolves at the rate at which its energy density redshifts relative to the total, and the coupled dark matter has the additional contribution from the mass drift. Although $\wV$ is not an independent variable, the same computation gives
\begin{equation}\label{eq:aut_V}
    \wV' = \wV\big(3(1+\weff) - c_1(\phi)\,u\big)\;,
\end{equation}
which has the same form as the fluid equations.

The physical region of the state space is the set
\begin{equation}
    \mathcal{S} = \big\{\, (u, \wDM, \wB, \wR) \;:\; \wDM \geq 0\,,\; \wB \geq 0\,,\; \wR \geq 0\,,\; \wV \geq 0 \,\big\}\;,
\end{equation}
where we assume that the potential is non-negative, $V(\phi) \geq 0$, and $\wV$ is understood as a function of the state variables through \cref{eq:constraint}. The condition $\wV \geq 0$ reads
\begin{equation}
    \frac{u^2}{6} + \wDM + \wB + \wR \leq 1\;,
\end{equation}
so every abundance lies in $[0,1]$ and $u^2 \leq 6$. The region $\mathcal{S}$ is therefore a closed and bounded subset of $\mathbb{R}^4$, hence compact. It is also preserved by the flow, in the sense that a trajectory which starts in $\mathcal{S}$ remains in $\mathcal{S}$ for all $N$. To see this, note that each of \cref{eq:aut_dm,eq:aut_b,eq:aut_r,eq:aut_V} has the form $\Omega_i' = \Omega_i f_i$ with $f_i$ a smooth function of the state variables. Along any trajectory this integrates to
\begin{equation}
    \Omega_i(N) = \Omega_i(0)\,\exp\bigg(\int_0^N f_i\,\d N'\bigg)\;,
\end{equation}
so the sign of each $\Omega_i$ is conserved. An abundance which is positive initially remains positive, and one which vanishes initially remains zero. Since $\mathcal{S}$ is precisely the set on which all four of $\wDM$, $\wB$, $\wR$, and $\wV$ are non-negative, no trajectory can leave it. Every trajectory therefore exists for all $N$, does not blow up, and has its limit points inside $\mathcal{S}$. This is what makes the classification of the following sections exhaustive.

\section{Exponential slopes and running slopes}\label{sec:theorem}

We now ask for which functions $V(\phi)$ and $M(\phi)$ the system \cref{eq:aut_sys_full} admits exact stasis. By exact stasis we mean a solution on which every abundance is constant in time. For a single fluid coupled to the scalar the answer is known. In the uncoupled case, exact scaling requires an exponential potential \cite{Liddle:1998xm}, and potentials which are asymptotically exponential approach the scaling solution \cite{Nunes:2000yc}. With a coupling to matter, the general Lagrangian admitting scaling solutions was derived for constant coupling in \cite{Piazza:2004df,Tsujikawa:2004dp}, and it was shown in \cite{Amendola:2006qi} that a field dependent coupling can always be absorbed into a field redefinition, so that a canonically normalized field must have constant coupling and an exponential potential. In this section we rederive this result directly from \cref{eq:aut_sys_full}. The derivation is short and does not depend on which components are present, and it shows that the classification of the next section, carried out for exponential $V$ and $M$, is in fact complete for arbitrary $V$ and $M$. We then determine what happens when the slopes are not constant. Treating a general potential as an exponential model with a slowly drifting slope is a standard device for an uncoupled scalar \cite{Steinhardt:1999nw,Ng:2001hs,Zhou:2007xp,Fang:2008fw}. Here we apply it to the coupled system and find that the outcome is fixed by an integrability condition on the running, which separates potentials which converge to stasis from potentials which track a drifting target indefinitely.

We must first be precise about what is held constant. The state variables of \cref{eq:aut_sys_full} are $u$, $\wDM$, $\wB$, and $\wR$, and the kinetic and potential fractions of the scalar are $u^2/6$ and $\wV$. We define exact stasis as a solution on which all of these are constant, so that the abundance of every component is constant with the scalar counted as two components. This is the same as a fixed point of \cref{eq:aut_sys_full}. If either spectator $\wB$ or $\wR$ is nonzero, this is equivalent to constancy of the fluid abundances alone, since the equation for the spectator pins $\weff$ to $0$ or $1/3$, and $\weff$ depends on the state only through $u^2$ once the abundances are fixed.\footnote{On the face $\wB = \wR = 0$ the two definitions differ. Constancy of $\wDM$ and $\wDE$ alone leaves the split of the dark energy between kinetic and potential energy free, and for any prescribed $u(N)$ of fixed sign one can construct a $V$ and $M$ along the trajectory which keep the abundances constant while $u$ varies. Such solutions are tuned to the trajectory, do not have power law expansion, and are destroyed by baryons or radiation, so we do not consider them further. For the exponential model of the following sections the two definitions agree on every face, and \cite{Amendola:2006qi} imposes constant $w_\phi$ for the same reason.}

With all state variables constant, $u$ is constant, and integrating $\phi' = u$ shows that the field moves along the straight trajectory
\begin{equation}\label{eq:straight}
    \phi(N) = \phi_0 + uN\;.
\end{equation}
Each of \cref{eq:aut_dm,eq:aut_b,eq:aut_r,eq:aut_V} has the form $\Omega_i' = \Omega_i \times (\text{rate})$, so each abundance either vanishes or forces its rate factor to vanish. For the coupled dark matter and for the potential energy, the rate factors involve the slopes, and along \cref{eq:straight} the conditions read
\begin{subequations}\label{eq:slope_conditions}
\begin{align}
    \wDM > 0 &\;\Rightarrow\; c_2(\phi_0 + uN)\,u = 3\,\weff\;,\label{eq:cprime_cond}\\
    \wV > 0 &\;\Rightarrow\; c_1(\phi_0 + uN)\,u = 3(1 + \weff)\;.\label{eq:c_cond}
\end{align}
\end{subequations}
The right hand sides are constants, since $u$ and $\weff$ are constant at a fixed point. The left hand sides are the functions $c_2(\phi)$ and $c_1(\phi)$ evaluated along a point which moves through field space. For $u \neq 0$ the point sweeps out an interval, so $c_2$ is constant on the traversed segment wherever dark matter is present, and $c_1$ is constant on the traversed segment wherever potential energy is present. Integrating \cref{eq:slopes}, $M$ is exponential on the segment in the first case and $V$ is exponential on the segment in the second.

Two caveats should be recorded. First, if $u = 0$ the field sits at a single value $\phi_0$, no interval is traversed, and \cref{eq:slope_conditions} constrain only the numbers $c_1(\phi_0)$ and $c_2(\phi_0)$. These are the solutions in which the field is frozen at an extremum of $V$ or at an extremum of $M$, since $u = 0$ with $\wDM > 0$ forces $\weff = 0$ while $u = 0$ with $\wV > 0$ forces $\weff = -1$, so the two cannot coexist, together with the points R and B where both vanish. The energy exchange between $\wDM$ and $\wDE$ is switched off, and they place no condition on the form of the potential. Second, where $\wV = 0$ there is no condition on $V$ at all. For a potential which is positive along the trajectory (so $\wV>0$) such a solution $\wV = 0$ is only approached asymptotically, and $\wV \rightarrow 0$ requires the rate factor in \cref{eq:aut_V} to be negative asymptotically. When the rate tends to a nonzero limit this reads
\begin{equation}\label{eq:V_ineq}
    c_1(\phi)\,u > 3(1+\weff)\;.
\end{equation}
The condition on $V$ in this case is therefore only an inequality requiring $V$ to be steep enough for its abundance to decay, rather than an equality fixing any parameters. A rate which tends to zero from below also removes $\wV$, only more slowly.

The same equations tell us what happens when the slopes are not constant. Since the fixed points depend on $V$ and $M$ only through $c_1$ and $c_2$, a potential with running slopes can be described as the exponential model with slowly drifting parameters. At each instant there is a fixed point of the exponential model with the current values $c_1(\phi(N))$ and $c_2(\phi(N))$, and the question is whether the trajectory follows it. For an uncoupled scalar and a single fluid this description was developed in \cite{Steinhardt:1999nw,Ng:2001hs}, where the fixed points at the instantaneous slope are called instantaneous critical points. It was cast as a dynamical system by adding the slope as a variable in \cite{Zhou:2007xp,Fang:2008fw}, with an application to an algebraically coupled fluid in \cite{Dutta:2017kch}. Here we apply it to the coupled system with two slopes, and we organize the outcomes by a single criterion. Suppose $c_1(\phi) \rightarrow c_{1,\infty}$ as $\phi \rightarrow \infty$ along the trajectory, with $c_2$ constant. The deviation of the slope from its limit acts as a source in \cref{eq:aut_u} which displaces the trajectory from the fixed point of the limiting exponential model, and the accumulated displacement is controlled by
\begin{equation}\label{eq:integrability}
    \int^\infty \big|c_1(\phi) - c_{1,\infty}\big|\,\d\phi\;.
\end{equation}
If this integral is finite, the solution converges to the exact stasis of the limiting exponential model. If the integral diverges, the trajectory never settles at a fixed point. It instead tracks the instantaneous fixed point,
\begin{equation}
    \wDM(N) \rightarrow \swDM\big(c_1(\phi(N)),\, c_2\big)\;,
\end{equation}
and approaches the limiting stasis only as a power law, at the rate set by the running. The convergent case for an uncoupled scalar was shown in \cite{Nunes:2000yc}, and an explicit example of asymptotic approach to a stasis attractor is studied in \cite{Barber:2026bse}.

The two cases correspond to two classes of potentials. A finite sum of exponentials,
\begin{equation}
    V(\phi) = \sum_i V_i\, e^{-\lambda_i \phi}\;,
\end{equation}
has a slope which approaches the smallest $\lambda_i$ exponentially fast, so the integral converges. This is the double exponential potential of \cite{Barreiro:1999zs} for example. A contribution which is a power of a modulus $R$ becomes an exponential in the canonically normalized field when the kinetic term of the modulus is logarithmic, since then $R \propto e^{k\phi}$, and a sum of powers becomes a sum of exponentials \cite{Copeland:2000vh}. In the DD every geometric contribution to the radion potential is of this type, and more generally it was argued in \cite{Etheredge:2026cha} that the asymptotic exponents of potential contributions in string theory take values on a lattice fixed by their scaling under Weyl rescalings and the string loop expansion. So is the uplift term $E/({\rm Re}\,T)^\alpha$ of \cite{Blanco-Pillado:2004aap} used for inflationary models. Nonperturbative contributions to the superpotential, such as gaugino condensates or instantons of the form $e^{-aT}$ in a modulus $T$, and the corrections to the Casimir energy from massive bulk fields which are of order $e^{-2\pi mR}$ \cite{Katayama:2026ras,Anchordoqui:2024gfa}, depend exponentially on the modulus itself. When the canonical field is logarithmic in the modulus, as for the radion and for moduli with $K \propto -\log(T + \bar{T})$, these become
\begin{equation}
    e^{-aR} = \exp\!\big(-a R_0\, e^{k\phi}\big)
\end{equation}
in the canonical field, multiplied by exponentials in $\phi$ from the power law prefactors. Such terms converge faster than any exponential and do not affect the asymptotic slope. For our purposes, these terms would matter only at moderate $\phi$ where they produce minima or hilltops. Racetrack potentials, whose superpotential contains two such terms \cite{Blanco-Pillado:2004aap}, and the dilaton potential from gaugino condensation \cite{Binetruy:1998rz}, are examples. For all of these potentials the question of whether stasis is reached reduces to the value of the asymptotic slope, which is how we treat it in \cref{sec:capture}.

The integral diverges instead for a polynomial prefactor,
\begin{equation}
    V(\phi) = f(\phi)\, e^{-c\phi}
\end{equation}
with $f$ a polynomial of degree $p$, which gives $c_1 = c - p/\phi + \mathcal{O}(\phi^{-2})$ at large $\phi$. Such prefactors arise from a logarithmic dependence on the modulus, since $\log R$ is linear in $\phi$. One loop terms of the Coleman-Weinberg form $m^4 \log(m^2/\mu^2)$ with $m \propto 1/R$, and the running of couplings with $R$ dependent thresholds, are of this kind.\footnote{The Casimir energy is the one loop Coleman-Weinberg potential of the KK tower, with one piece removed. The removed piece is the vacuum energy of the uncompactified theory, which depends on $R$ only through the volume and is absorbed into the five dimensional cosmological constant. What remains comes from loops which wrap the circle. It is finite and it is suppressed as $e^{-2\pi mR}$ for a bulk field of mass $m \gg 1/R$, so only fields lighter than the KK scale contribute \cite{Arkani-Hamed:2007ryu}. The logarithms discussed here arise instead from four dimensional states whose mass depends on $R$. In the DD the masses of the SM fields on the brane typically would not depend on $R$, with the exception of the active neutrinos whose masses arise from a seesaw with a bulk right handed tower and scale with the species scale \cite{Montero:2022prj}.} The potential of \cite{Albrecht:1999rm}, with $f = A + (\phi - B)^2$, is the standard phenomenological example. The Casimir energies of \cite{Katayama:2026ras} contain no such logarithms, as expected in an odd number of dimensions, and we are not aware of a computed source of them for the DD radion. A candidate is the one loop vacuum energy of the active neutrinos, whose masses can depend on $R$ through the seesaw with the bulk tower \cite{Montero:2022prj}, which would give a term of the form $m_\nu(R)^4 \log m_\nu(R)^2$ and hence a non-integrable tail. We therefore regard the polynomial prefactor as the generic way in which such a tail can arise, and as a possibility in the DD scenario rather than an established prediction of it.

\begin{figure}[htbp]
    \centering
    \includegraphics[width=0.85\linewidth]{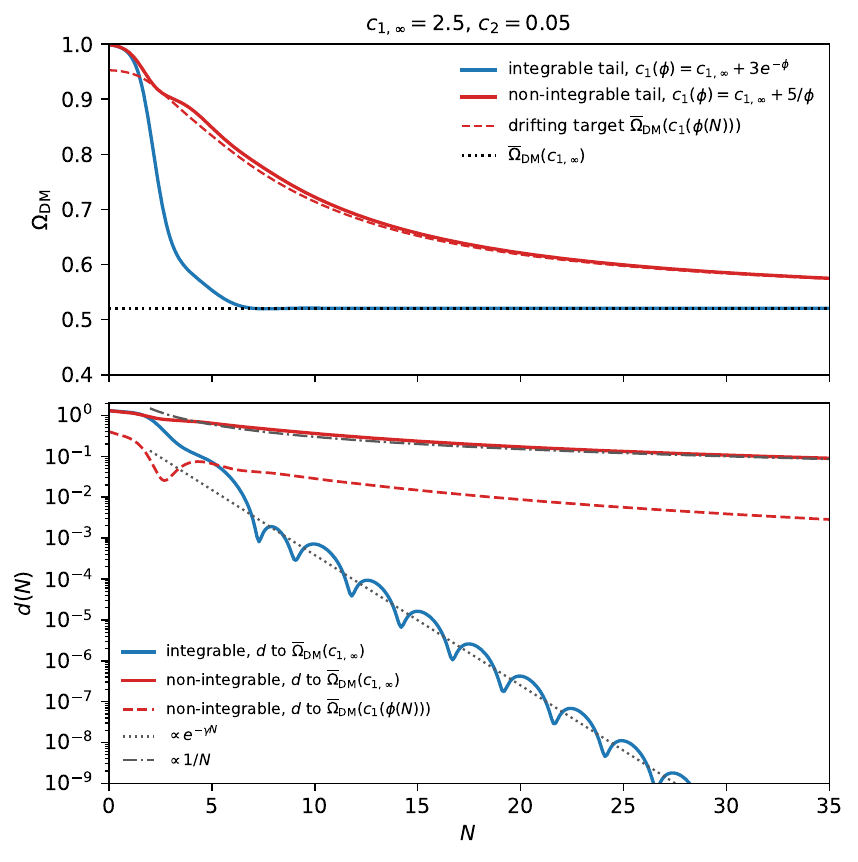}
    \caption{The two cases of \cref{eq:integrability}, for $c_{1,\infty} = 2.5$ and
    $c_2 = 0.05$. Top, the dark matter abundance for an integrable running of the
    slope (blue) and a non-integrable running (red), together with the drifting
    target $\swDM(c_1(\phi(N)))$ (dashed) and the stasis value of the limiting
    exponential model (dotted). Bottom, on a logarithmic scale, the distance
    $d(N)$ in the $(u, \wDM)$ plane between the trajectory and the fixed point.
    The integrable case converges to the asymptotic fixed point as $e^{-\gamma N}$,
    with $\gamma$ given by \cref{eq:gamma_def}. The
    non-integrable case approaches the asymptotic fixed point only as $1/N$, while
    its distance to the drifting target is roughly an order of magnitude smaller.}
    \label{fig:running_tails}
\end{figure}

The two cases are shown in \cref{fig:running_tails}. For a slope which approaches its limit exponentially, the trajectory converges to the stasis of the limiting exponential model, and the distance to it decreases as $e^{-\gamma N}$ with $\gamma$ the convergence rate of that model. For a slope which approaches its limit as $1/\phi$, the trajectory never converges. It follows the fixed point at the instantaneous value of the slope, staying roughly an order of magnitude closer to this moving target than to the asymptotic value, and approaching the latter only as $1/N$. The convergence rate $\gamma$ and a method for computing the drift are given in \cref{sec:S_explicit} and \cref{app:linresp}.

\section{Constant abundance solutions of the exponential model}\label{sec:classification}

By the results of \cref{sec:theorem}, exact stasis requires
\begin{equation}\label{eq:exp_forms}
    V(\phi) = V_0\,e^{-c_1\phi}\;,\qquad M(\phi) = M_0\,e^{-c_2\phi}\;,
\end{equation}
with constant $c_1$ and $c_2$. We take $c_2 > 0$ by the convention of \cref{sec:setup}, and both signs of $c_1$ are physical. For the DD, $c_2 > 0$ with $\phi \rightarrow +\infty$ is the decompactification limit with a light tower of KK modes of mass scale $M \sim 1/R$.

The critical points of \cref{eq:aut_sys_full} with \cref{eq:exp_forms} are the exact stasis solutions. They can be enumerated completely using the product structure of the fluid equations. Each abundance either vanishes or pins one linear condition on $\weff$ and $u$, and the remaining equation $u' = 0$ together with the constraint then fixes the rest. By the argument of \cref{sec:theorem}, the list obtained in this way is also the complete list of fixed points for arbitrary $V$ and $M$. This structure of critical points is not new. It was derived in \cite{Amendola:1999er,Amendola:2001rc} for exactly this four component cosmology, building on \cite{Wetterich:1994bg,Copeland:1997et,Ferreira:1997hj}. The dictionary between our conventions and those of \cite{Amendola:2001rc} is
\begin{equation}
    \mu = \sqrt{\tfrac{3}{2}}\,c_1\;,\qquad \beta = -\sqrt{\tfrac{3}{2}}\,c_2\;,
\end{equation}
and with this dictionary our \cref{tab:critpoints} reproduces the tables of \cite{Amendola:2001rc}.\footnote{The points $a$, $b_r$, $b_c$, $b_b$, $c_r$, $c_{rc}$, $c_c$, $d$ and $e$, and $f_b$ correspond to V, RS, S, BS, R, KRD, $\phi$MDE, K$_\pm$, and B respectively.} We rederive the classification here in our conventions for completeness. Our interest is in the aspects which the earlier literature did not pursue, namely the identification of the solutions as cosmological stasis, their lifetimes and exits in \cref{sec:lifetimes}, and their observational status in \cref{sec:obs}.

\begin{table}[htbp]
    \centering
    \begin{tabular}{ l  l  l  l  l  l }
    \toprule
    & point & $(u,\ \wDM,\ \wB,\ \wR)$ & $\weff$ & exists for \\
    \midrule
    R & radiation dom. & $(0,0,0,1)$ & $1/3$ & always \\
    B & matter dom.\ (Einstein-de Sitter) & $(0,0,1,0)$ & $0$ & always \\
    K$_\pm$ & kination & $(\pm\sqrt{6},0,0,0)$ & $1$ & always \\
    $\phi$MDE & DM + kinetic & $(2c_2,\ 1-\tfrac{2c_2^2}{3},\ 0,\ 0)$ & $\tfrac{2c_2^2}{3}$ & $c_2^2 \leq \tfrac{3}{2}$ \\
    KRD & kination + rad.\ + DM & $(\tfrac{1}{c_2},\ \tfrac{1}{3c_2^2},\ 0,\ 1-\tfrac{1}{2c_2^2})$ & $1/3$ & $c_2 \geq \tfrac{1}{\sqrt{2}}$ \\
    V & scalar dom. & $(c_1,0,0,0)$ & $\tfrac{c_1^2}{3}-1$ & $c_1^2 \leq 6$ \\
    RS & radiation scaling & $(\tfrac{4}{c_1},0,0,1-\tfrac{4}{c_1^2})$ & $1/3$ & $c_1^2 \geq 4$ \\
    BS & baryon scaling & $(\tfrac{3}{c_1},0,1-\tfrac{3}{c_1^2},0)$ & $0$ & $c_1^2 \geq 3$ \\
    S & coupled DM and DE stasis & $(\tfrac{3}{c_1-c_2},\ \swDM,0,0)$ & $\tfrac{c_2}{c_1-c_2}$ & \cref{eq:stasis_V0_pos} \\
    \bottomrule
    \end{tabular}
    \caption{The constant abundance solutions of the four component cosmology for
    $c_2 > 0$ and $V_0 > 0$, rederived from \cite{Amendola:1999er,Amendola:2001rc}.
    $\swDM$ is given in \cref{eq:swDM_value}. The points R, B, K$_\pm$, $\phi$MDE, and KRD lie
    on the invariant face $V = 0$. The coupling between dark matter and dark energy is active at $\phi$MDE, KRD, and S.}
    \label{tab:critpoints}
\end{table}

\begin{table}[htbp]
    \centering\footnotesize
    \setlength{\tabcolsep}{4pt}
    \begin{tabular}{ l  l  c  c  c  c  c }
    \toprule
    point & attractor when & $u$ & $\wDM$ & $\wB$ & $\wR$ & $\wV$ \\
    \midrule
    R & never & $-$ & $+$ & $+$ & $\ast$ & $+$ \\
    B & never & $-$ & $0$ & $\ast$ & $-$ & $+$ \\
    K$_+$ & never & $c_1 < \sqrt{6}$ & $c_2 < \sqrt{3/2}$ & $+$ & $+$ & $c_1 < \sqrt{6}$ \\
    K$_-$ & never & $c_1 > -\sqrt{6}$ & $+$ & $+$ & $+$ & $c_1 > -\sqrt{6}$ \\
    $\phi$MDE & never & $-$ & $\ast$ & $+$ & $c_2 > 1/\sqrt{2}$ & $c_1 < c_2 + \tfrac{3}{2c_2}$ \\
    KRD & never & $-$ & $\ast$ & $+$ & $\ast$ & $c_1 < 4c_2$ \\
    V & $c_1^2 < 3$, $c_1(c_1 - c_2) < 3$ & $-$ & $c_1(c_1 - c_2) > 3$ & $c_1^2 > 3$ & $c_1^2 > 4$ & $\ast$ \\
    RS & never & $-$ & $c_1 > 4c_2$ or $c_1 < 0$ & $+$ & $\ast$ & $\ast$ \\
    BS & $c_1 > 0$ & $-$ & $c_1 < 0$ & $\ast$ & $-$ & $\ast$ \\
    S & $c_1 < 0$ & $-$ & $\ast$ & $c_1 > 0$ & $c_2 < c_1 < 4c_2$ & $\ast$ \\
    \bottomrule
    \end{tabular}
    \caption{Stability of the critical points of \cref{tab:critpoints} for $c_2 > 0$,
    within their ranges of existence. The second column gives the condition under which
    the point is an attractor of the full system. The remaining columns give, for each
    state variable, whether a perturbation in that direction grows, with $+$ for always, $-$ for never,
    $0$ marginal, a condition when it grows only for those parameters, and $\ast$ when
    the component is present at the point so that its perturbation is not an independent
    direction. These follow from the signs of the eigenvalues of the Jacobian and the
    components of the corresponding eigenvectors. Perturbations of an absent fluid grow
    at the rate \cref{eq:spectator_lambdas}, and a growing $\wV$ perturbation is a transfer
    of energy from the present fluids into the potential. A point with no growing direction
    is an attractor, one with every direction growing is a repeller, which happens for
    K$_\pm$ when all their conditions hold, and every other case is a saddle. B is
    non-hyperbolic. The approach along the stable directions is oscillatory in parts of
    parameter space for KRD, RS, BS, and S, see \cite{Amendola:2001rc} for the full spectrum of
    eigenvalues.}
    \label{tab:stability}
\end{table}

The stability of each point is determined by linearizing \cref{eq:aut_sys_full} about the point. Writing the state as $X = X_* + \delta X$, small perturbations obey
\begin{equation}
    \delta X' = J\,\delta X\;,
\end{equation}
with $J$ the $4\times4$ Jacobian evaluated at $X_*$, so that $\delta X \sim e^{\lambda N}$ along the eigendirections. A point is an attractor if all eigenvalues have negative real part, a repeller if all have positive real part, and a saddle otherwise. Complex pairs indicate a spiral. The product structure again does most of the work. At any point where a component vanishes, the corresponding row of $J$ has a single entry on the diagonal, so each absent component contributes its own rate factor as an eigenvalue,
\begin{equation}\label{eq:spectator_lambdas}
    \lambda_i = 3\big(\weff^{(*)} - w_i\big) - \delta_{i,\m{DM}}\,c_2u_*\;,\qquad
    \lambda_V = 3\big(1 + \weff^{(*)}\big) - c_1\,u_*\;,
\end{equation}
where stars indicate the quantity is evaluated at the relevant critical point. These are the rates at which a small admixture of the absent component grows or decays against the background solution. The remaining block of $J$ is at most $2\times2$. The resulting classification is collected in \cref{tab:stability}. We list the character of each point and the directions along which perturbations grow rather than the eigenvalues themselves, which are given in \cite{Amendola:2001rc}. Any eigenvalues which enter the later analysis are quoted where they are used.

The dependence of the eigenvalues on the parameters encodes a sequence of stability exchanges. The scalar dominated point V is the attractor when $c_1^2 < 3$ and $c_1(c_1 - c_2) < 3$. On the branch $c_1 > 0$ the second condition follows from the first, while on the branch $c_1 < 0$ it is the stronger one. The dark matter eigenvalue of V, $c_1^2 - c_1c_2 - 3$, changes sign on the boundary
\begin{equation}
    c_1(c_1 - c_2) = 3
\end{equation}
of the stasis region. On the branch $c_1 < 0$ this is where V hands stability to S. On the branch $c_1 > 0$ the baryon eigenvalue $c_1^2 - 3$ changes sign first, and V hands stability to BS at $c_1^2 = 3$. The radiation scaling solution RS acquires an unstable dark matter direction for $c_1 > 4c_2$, which is exactly where KRD becomes stable in its $V$ direction. On the line $c_1 = 4c_2$ a degenerate one parameter family of solutions interpolates between RS and KRD. Finally BS is an attractor for $c_1 > 0$ but is unstable for $c_1 < 0$, where the coupled dark matter grows against it.

\subsection{The coupled stasis S}\label{sec:S_explicit}
\begin{figure}[htbp]
    \centering
    \includegraphics[width=0.49\linewidth]{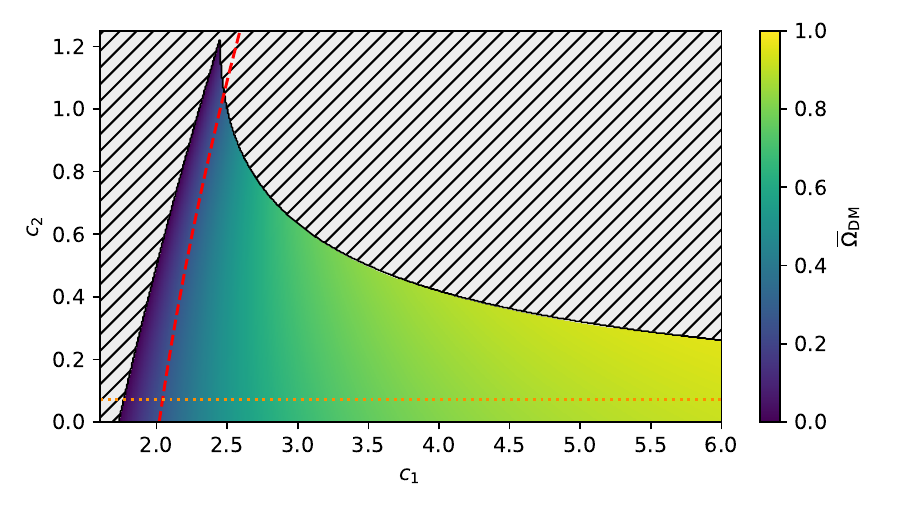}\hfill
    \includegraphics[width=0.49\linewidth]{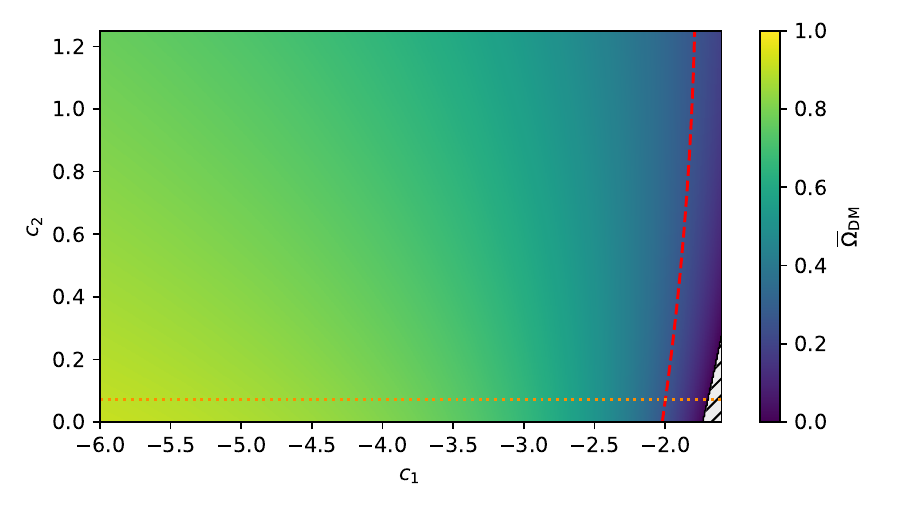}
    \caption{The stasis value $\swDM$ of \cref{eq:swDM_value} on the branch $c_1 > 0$
    (left) and the branch $c_1 < 0$ (right). The red dashed curve marks $\swDM = 0.264$,
    the present dark matter fraction $\Omega_{\m{c}} = \Omega_{\m{c}}h^2/h^2$ of
    \cite{Planck:2018vyg}. The orange dotted line is the dark fifth-force
    bound $c_2 = 0.071$ derived in \cref{sec:combined} from \cite{Archidiacono:2022iuu}, and the region above it is excluded. The hatched regions do not admit stasis solutions.}
    \label{fig:region_maps}
\end{figure}

The point S is the stasis of central interest. It is the only solution with both dark components at nonvanishing abundance and the coupling active. We record its construction in cosmic time, which also determines the normalization of the potential. On the face $\wB = \wR = 0$, \cref{eq:wDM_eom} says that stasis requires
\begin{equation}\label{eq:stasis_cond}
    \dot{\phi}\,\frac{\d\log M}{\d\phi} + 3\,\sweff H = 0\;.
\end{equation}
During stasis the Hubble parameter evolves as
\begin{equation}
    H = \frac{\overline{\kappa}}{3t}\;,\qquad \overline{\kappa} = \frac{2}{1+\sweff}\;,
\end{equation}
and the scale factor evolves as $a \propto t^{\overline{\kappa}/3}$. The exponential forms admit the solution $\phi(t) = \alpha\log t + \beta$ with
\begin{equation}
    \alpha = \frac{2}{c_1}\;,\qquad \overline{\kappa} = 2\Big(1 - \frac{c_2}{c_1}\Big)\;,\qquad
    e^{-c_1\beta} = \frac{6 + 4c_2(c_2-c_1)}{3c_1^2 V_0}\;,
\end{equation}
and the stasis values are
\begin{equation}\label{eq:swDM_value}
    \swDM = \frac{c_1(c_1-c_2) - 3}{(c_1-c_2)^2}\;,\qquad
    u_s = \frac{3}{c_1-c_2}\;,\qquad
    \sweff = \frac{c_2}{c_1-c_2}\;.
\end{equation}
One can check that
\begin{equation}
    w_\phi = \frac{c_2(c_1-c_2)}{3 - c_2(c_1-c_2)}
\end{equation}
is constant, as required. Restricting to $e^{-c_1\beta} > 0$ and $0 < \swDM < 1$ with $V_0 > 0$, we find the following conditions on $c_1$ and $c_2$ for stasis,
\begin{equation}\label{eq:stasis_V0_pos}
    c_1(c_1-c_2) - 3 > 0\;,\qquad 2c_2(c_1-c_2) - 3 < 0\;.
\end{equation}
If instead $V_0 < 0$ is allowed, one finds stasis solutions in the region $0 < 2c_2(c_1-c_2) - 3 < 3$, but these are unstable. Note that the conditions \cref{eq:stasis_V0_pos} are satisfied on two branches. On the branch $c_1 > 0$ the first condition requires $c_1 \gtrsim \sqrt{3}$ at small $c_2$. On the branch $c_1 < 0$ the first condition reads
\begin{equation}
    |c_1|\big(|c_1| + c_2\big) > 3\;,
\end{equation}
and the second condition is automatic. \Cref{fig:region_maps} shows $\swDM$ on both branches. On the branch $c_1 > 0$ one has $\sweff > 0$, while on the branch $c_1 < 0$ one has $\sweff < 0$. By \cref{eq:spectator_lambdas} this sign determines the fate of each branch in the presence of baryons, as we discuss in the next section.

Within the face $\wB = \wR = 0$, the $2\times2$ block of the Jacobian at S has
\begin{equation}\label{eq:trdet_fp}
    \Tr J = -\frac{3}{2}\,\frac{c_1-2c_2}{c_1-c_2}\;,\qquad
    \det J = \frac{3}{2}\,\swDM\big(3 - 2c_2(c_1-c_2)\big)\;,
\end{equation}
with eigenvalues
\begin{equation}\label{eq:eigenvalues}
    \lambda_\pm = \frac{1}{2}\Big(\Tr J \pm \sqrt{(\Tr J)^2 - 4\det J}\Big)\;.
\end{equation}
One can verify that stability within the face, which requires $\det J > 0$ and $\Tr J < 0$, is equivalent to the conditions \cref{eq:stasis_V0_pos}. The eigenvalues also set the rate of convergence toward stasis. Perturbations decay as $e^{-\gamma N}$ with
\begin{equation}\label{eq:gamma_def}
    \gamma = -\max\mathrm{Re}(\lambda_\pm)\;,
\end{equation}
which for spirals is $\gamma = \frac{3}{4}(c_1-2c_2)/(c_1-c_2)$. In the decoupled limit $c_2 \rightarrow 0$, the point S reduces to the scaling solution of exponential quintessence and \cref{eq:eigenvalues} reproduces the eigenvalues
\begin{equation}
    \lambda_\pm = -\frac{3}{4}\bigg[1 \mp \sqrt{\frac{24}{c_1^2} - 7}\,\bigg]
\end{equation}
of \cite{Copeland:1997et}.

\section{Coupled stasis S lifetime and exit}\label{sec:lifetimes}

Viewed as cosmological stasis in the sense of \cite{Dienes:2021woi,Dienes:2023ubz}, the solutions of \cref{tab:critpoints} are epochs, and one would like to know how long they last and what ends them. This information is determined by the transverse eigenvalues \cref{eq:spectator_lambdas}. Here we will focus on analyzing the lifetime of the coupled stasis S. At the coupled stasis S, the transverse eigenvalues are
\begin{equation}\label{eq:transverse_S}
    \lb = 3\sweff = \frac{3c_2}{c_1-c_2}\;,\qquad
    \lr = 3\sweff - 1 = \frac{4 c_2 - c_1}{c_1 - c_2}\;.
\end{equation}
Uncoupled dust and radiation scales against any stasis background as
\begin{equation}
    \wB \propto a^{3\sweff} = a^{\lb}\;,\qquad \wR \propto a^{3\sweff - 1} = a^{\lr}\;.
\end{equation}

First, consider the branch $c_1 < 0$  where one has $\sweff < 0$. In this case, since $c_2 > 0$ it is clear that both $\lb < 0$ and $\lr < 0$, so both spectators decay. Here, S is an attractor of the full system and stasis is eternal. How soon stasis could be reached in this branch from the present state of the universe is computed in \cref{sec:anchoring}.

\begin{figure}[htbp]
    \centering
    \includegraphics[width=\linewidth]{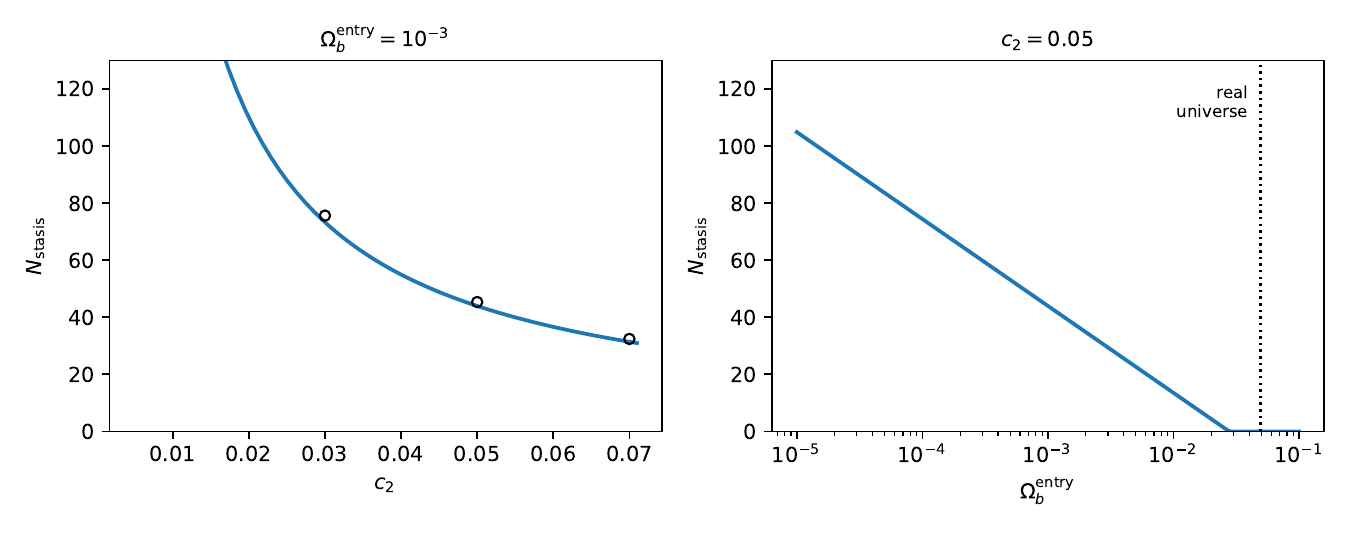}
    \caption{The quasi-stasis lifetime \cref{eq:N_stasis} on the branch $c_1 > 0$ with
    $\swDM = 0.264$ and $\delta = 0.1$. The left plot shows the lifetime as a function of the coupling at
    $\wB^{\m{entry}} = 10^{-3}$. Open markers are lifetimes measured by direct
    integration of \cref{eq:aut_sys_full} with the same tolerance. The right plot shows the lifetime as a function
    of the entry baryon abundance at $c_2 = 0.05$. The dotted line marks the baryon
    abundance with which the stasis is entered in our universe, which lies beyond the
    point where the stasis lifetime vanishes.}
    \label{fig:lifetimes}
\end{figure}

Now consider the branch where $c_1 > 0$. Existence of the stasis solution requires $c_1 > c_2$ (recall \cref{eq:stasis_V0_pos}), so $\lb > 0$. However, the sign of $\lr$ is not fixed by existence alone. From $\lr = (4c_2 - c_1)/(c_1 - c_2)$, radiation decays in the stasis solution when $c_1 > 4c_2$, i.e. when $\sweff < 1/3$. The lower boundary in $c_1$ of the stasis region lies above $4c_2$ whenever $c_2 < 1/2$, in which case $\lr < 0$ throughout the region. The dark fifth-force constraints of \cref{sec:combined} require $c_2$ to be far below $1/2$ today, so for the modeling of the present universe radiation always decays in the stasis solution. The baryons therefore grow, and S is a saddle with exactly one unstable direction. We note that this constraint applies to the coupling as measured today. A coupling which varies over field space could have been larger in the early universe, or become larger in the future, and we return to this possibility in \cref{sec:capture}. The baryon growth rate is slow, with $\lb \lesssim 0.13$ for couplings allowed by fifth-force bounds, while the convergence rate within the dark sector is $\gamma \simeq 3/4$. Trajectories are therefore drawn in, linger, and depart. To make the lifetime of this epoch precise we define $N_{\m{stasis}}$ as the number of e-folds during which every abundance stays within a fraction $\delta$ of its stasis value. During the epoch the trajectory tracks the drifting quasi-stasis line
\begin{equation}\label{eq:qstasis_line}
    u = u_s + k_u\,\wB\;,\qquad \wDM = \swDM + k_\Omega\,\wB\;,\qquad \wB \propto e^{\lb N}\;,
\end{equation}
with coefficients $k_u$ and $k_\Omega$ derived in \cref{app:linresp}. We find $k_\Omega \simeq -1$ across the parameter space, so the baryons grow at the expense of the dark matter while the total dust abundance stays near $\swDM$. On the quasi-stasis line $\weff = \sweff + \mathcal{O}(\wB)$, so \cref{eq:aut_b} integrates to
\begin{equation}
    \wB(N) = \wB^{\m{entry}}\, e^{\lb N}
\end{equation}
at leading order, with $N$ counted from the entry into stasis and $\wB^{\m{entry}}$ the baryon abundance at that time. The second relation in \cref{eq:qstasis_line} then gives the fractional deviation of the dark matter from its stasis value as $|k_\Omega|\,\wB(N)/\swDM$, which grows exponentially at the rate $\lb$. The dark matter leaves its tolerance band when this deviation reaches $\delta$, that is when $|k_\Omega|\,\wB^{\m{entry}} e^{\lb N} = \delta\swDM$, so that
\begin{equation}\label{eq:N_stasis}
    N_{\m{stasis}} = \frac{1}{\lb}\,\ln\frac{\delta\,\swDM}{|k_\Omega|\,\wB^{\m{entry}}}\;,
\end{equation}
if the argument of the logarithm exceeds one, and $N_{\m{stasis}} = 0$ otherwise. We use the dark matter band to define $N_{\m{stasis}}$ here rather than the band on $u$ because $k_u$ is proportional to the small coupling $c_2$ while $k_\Omega \simeq -1$, so the dark matter always leaves its band first. The expression is the leading term in an expansion in $\wB$, and it agrees with direct integration of \cref{eq:aut_sys_full} to within a few percent. The exact lifetimes are slightly longer, see \cref{fig:lifetimes}. For $\swDM = 0.264$, $c_2 = 0.05$, $\delta = 0.1$, and $\wB^{\m{entry}} = 10^{-3}$ it gives $44$ e-folds, and the lifetime grows as $1/c_2$ toward weak coupling.

However, the entry value $\Omega_{\m{b}}^{\m{entry}}$ is not a free parameter in a realistic model. Baryons make up a fraction $F_{\m{b}} = 0.157$ of the matter in our universe \cite{Planck:2018vyg}, so a stasis with $\swDM = 0.264$ is entered with $\wB^{\m{entry}} \simeq 0.05$. Then $|k_\Omega|\wB^{\m{entry}}$ already exceeds $\delta\swDM$ for $\delta = 0.1$ and as a result $N_{\m{stasis}} = 0$. The same holds for every $\swDM$ and every allowed $c_2$, as the right panel of \cref{fig:lifetimes} shows, since $N_{\m{stasis}}$ falls linearly in $\log\wB^{\m{entry}}$ with slope $1/\lb$ and reaches zero before the value of our universe. The baryons are already too abundant when the dark sector arrives, and the drift toward the baryon scaling solution BS begins immediately. A lifetime of tens of e-folds requires $\wB^{\m{entry}} \lesssim 10^{-3}$, which means a stasis reached long before the baryons were a comparable fraction of the matter, and in \cref{sec:obs} we will show that such an early stasis is excluded. After the exit the trajectory flows to BS, which is the attractor of the branch, and the coupled dark matter fades away as the radion continues to roll.

\section{Observational viability}\label{sec:obs}

A stasis attractor with $\swDM \approx 0.26$ is exactly the kind of structure which could explain why $\wDM$ and $\wDE$ are comparable today, since the abundances would be fixed by dynamics rather than by initial conditions. In this section we determine whether the observed universe allows this. We will find that the universe cannot have reached stasis in the past, is not in stasis now, and for physically motivated potentials will not be captured by stasis in the future. Each of these statements is derived below together with the assumptions under which it could fail.

\subsection{Fifth-force constraints}\label{sec:combined}
We begin with the constraint which restricts the coupling $c_2$ from the outset. In the model under study the radion is not stabilized and remains effectively massless. The dark sector therefore experiences an attractive force due to the exchange of $\phi$, as noted in \cite{Bedroya2025}. The ratio of this force to gravity is
\begin{equation}
    \frac{F_\phi}{F_{\m{grav}}} = 2c_2^2\;.
\end{equation}
Data from the cosmic microwave background (CMB) and baryon acoustic oscillations (BAO) bound the strength of a fifth-force in the dark sector to less than a percent of the strength of gravity, specifically $F_\phi/F_{\m{grav}} < 0.0053$ \cite{Archidiacono:2022iuu}. This bound weakens when the light scalar also acts as the dark energy. We will use the conservative bound $F_\phi/F_{\m{grav}} < 10^{-2}$, which gives
\begin{equation}\label{eq:fifthforce}
    |c_2| \lesssim 0.071\;.
\end{equation}
The bound $c_2 \lesssim 0.2$ from tidal streams quoted in \cite{Bedroya2025} is weaker. These bounds constrain the coupling as measured today, but the early coupling is not unconstrained either. A coupling active before recombination enhances the self gravity of dark matter by a factor $1 + 2c_2^2$ while leaving baryons untouched, which alters the dark matter growth factor and the baryon to dark matter ratio at decoupling and shifts the acoustic peaks \cite{Pettorino:2012ts,Baldi:2010vv}. CMB data alone bound a constant coupling to $c_2 < 0.066$ at $95\%$ confidence \cite{Planck:2015bue}, so a coupling of order one during the CMB era is excluded even if it is small today. A coupling which is large only well before recombination or only in the future remains open, and to our knowledge no likelihood analysis of field dependent couplings exists.

\subsection{Stasis in the early universe}\label{sec:early}

Between BBN and matter-radiation equality the expansion is radiation dominated, so a stasis epoch in this window must contain radiation. The solutions with $\wR = 0$ could only have occurred earlier, and since the model we consider contains no source of radiation and no reheating, an epoch of one of them would have to be followed by the standard radiation era through dynamics outside the current model we study. We will not pursue studying such models here. 

The stasis solutions which contain radiation are RS and KRD. RS is the scaling solution of an uncoupled exponential potential and carries a scalar fraction $\Omega_\phi = 4/c_1^2$ with no dark matter. KRD is a stasis of dark matter, radiation, and kinetic energy with $\weff = 1/3$ exactly, so that the background expansion is that of radiation domination. It is sustained by the mass drift, which pumps energy into the kinetic term. However, KRD cannot be an epoch of our cosmological history if the slopes are constant. Its share of energy outside the SM is
\begin{equation}
    \Omega_{\m{kin}} + \wDM = \frac{1}{2c_2^2} \geq \frac{1}{3}
\end{equation}
for $c_2 \leq \sqrt{3/2}$, which violates the bound $\Omega_\phi(T \sim \m{MeV}) \lesssim 0.045$ on a non-radiation component at BBN \cite{Bean:2001wt}, see also \cite{Fields:2019pfx}, if it persists to that time. Its baryon eigenvalue is $+1$, so the baryons grow against it and end it, but the exit does not lead to a standard cosmology. Once the field rolls with constant slope in the radiation era, the attractor of the exponential potential is the scaling solution RS for $c_1^2 > 4$ and the scalar dominated point V for $c_1^2 < 4$. V is excluded, and RS retains $\Omega_\phi = 4/c_1^2$, which satisfies the BBN bound only for $|c_1| > 9.4$. If the baryons instead overtake the radiation first, the trajectory reaches BS, where $u = 3/c_1$ and, on the branch $c_1 > 0$, the coupled dark matter redshifts faster than the baryons and is removed. On the branch $c_1 < 0$ the dark matter grows at BS and the flow continues to S. Standard early cosmology therefore requires the scalar to remain frozen and away from all of these attractors until late times. The anchored cosmologies of the next subsection arrive at the same conclusion from the opposite direction. KRD therefore has an exit, but no exit into the standard cosmology, and we return to this point in \cref{sec:production} and \cref{sec:conc}.

\subsection{Anchored cosmologies}\label{sec:anchoring}
The analysis is made sharp by an anchoring construction which removes all freedom from the model at fixed $c_1$ and $c_2$. At matter-radiation equality every fluid abundance is fixed by measurement. There, $\wR = \Omega_{\m{m}}$ by the definition of equality, the baryon fraction of matter is
\begin{equation}
    \frac{\wB}{\Omega_{\m{m}}} = \frac{\Omega_{\m{b}}h^2}{\Omega_{\m{b}}h^2 + \Omega_{\m{c}}h^2} \simeq 0.157\;,
\end{equation}
and the coupled dark matter is the remainder. Throughout we use the Planck 2018 values from the combination of temperature, polarization, and lensing data \cite{Planck:2018vyg}, namely $\Omega_{\m{b}}h^2 = 0.02237$, $\Omega_{\m{c}}h^2 = 0.1200$, $\wDE^{(0)} = 0.685$, and $z_{\m{eq}} = 3402$. The scalar is frozen at this time by Hubble friction, as also noted in \cite{Bedroya2025}. This leaves a single number, the frozen energy fraction of the scalar,
\begin{equation}
    \varepsilon = \wV(z_{\m{eq}}) = \frac{V(\phi_i)}{3H^2(z_{\m{eq}})}\;,
\end{equation}
which we fix by requiring that the solution reproduces the measured value $\wDE = 0.685$ today \cite{Planck:2018vyg}. After anchoring, every late time observable is a prediction of $c_1$ and $c_2$ with no free parameters.

It is important to be careful about what quantities we compare to observational data. The dynamics and the stasis solutions are formulated for the physical components, the tower with energy density $\rho_{\m{DM}} = nM(\phi)$ and the radion with $\rho_\phi$. Fits of $\Lambda$CDM or of a $w_0 w_a$ parameterization instead take matter to be the component which scales as $a^{-3}$, normalized to its value today \cite{Das:2005yj}, and dark energy to be the remainder of the Friedmann equation, \begin{equation}\label{eq:eff_split}
    \rho_{\m{DE}}^{\m{eff}}(a) = 3H^2 - \rho_{\m{m},0}\,a^{-3} - \rho_{\m{r}}(a) = \rho_\phi + \rho_{\m{DM},0}\,a^{-3}\Big(e^{-c_2[\phi(a) - \phi_0]} - 1\Big)\;,
\end{equation}
with the corresponding equation of state $w_{\m{DE}}^{\m{eff}} = -1 - \tfrac{1}{3}\,\d\ln\rho_{\m{DE}}^{\m{eff}}/\d N$. The split does not affect the anchoring. At $a = 1$ the effective and physical matter densities coincide by construction, so the condition $\wDE = 0.685$ is the same in both descriptions. At matter-radiation equality the CMB measures the physical ratio $\rho_{\m{m}}/\rho_{\m{r}}$ at recombination directly, with the field frozen and $M$ constant in between, so the physical equality redshift and baryon fraction coincide with the $\Lambda$CDM values.\footnote{The $a^{-3}$ extrapolation of $\Omega_{\m{c}}h^2$ from today is therefore never used, and the matter density today is an output of the model rather than an input.} The anchored trajectories are the same whichever split is used. Of the quantities we compare to data, the total equation of state $\weff = -1 - \tfrac{2}{3}\,\dot{H}/H^2$ is independent of any split, and only $w_{\m{DE}}^{\m{eff}}(z)$ of \cref{sec:accel} depends on it.

Note that in a stasis epoch the physical abundances are constant while $\rho_{\m{m},0}a^{-3}/3H^2$ is not, so an observer fitting $\Lambda$CDM to a universe in stasis would infer an evolving dark energy even though nothing in the dark sector is changing. A full likelihood comparison of the model with the data, in which the $\Lambda$CDM inferred parameters themselves shift, is beyond the scope of this paper and has been carried out for the DD in \cite{Bedroya2025}.

\begin{figure}[htbp]
    \centering
    \includegraphics[width=0.8\linewidth]{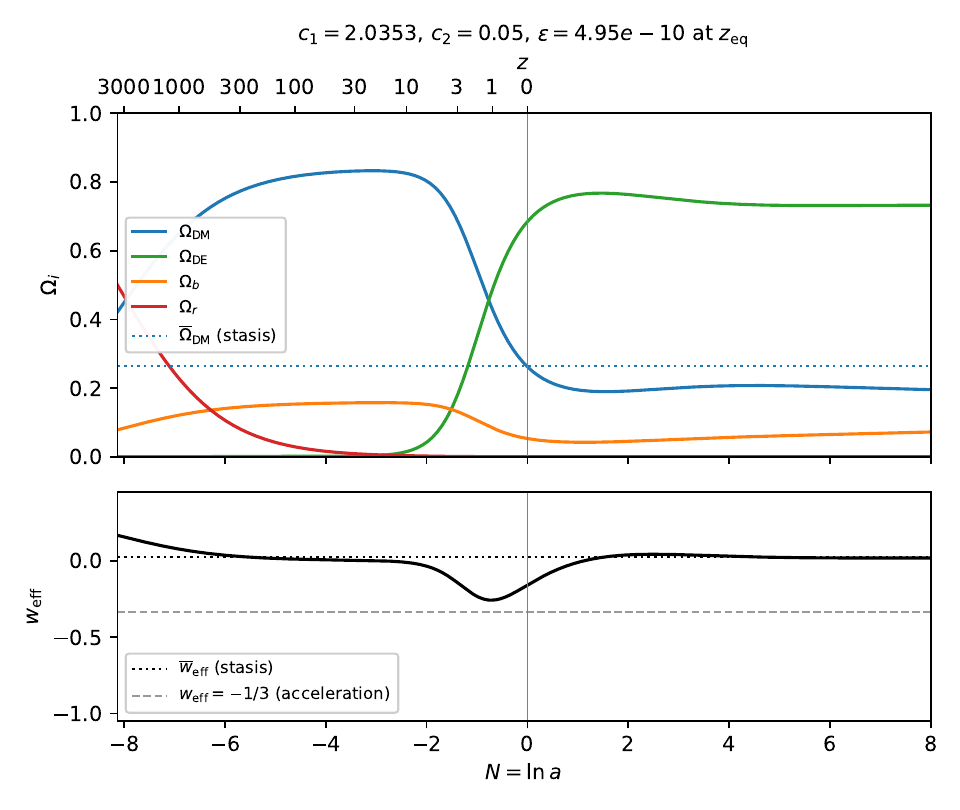}\\
    \includegraphics[width=0.8\linewidth]{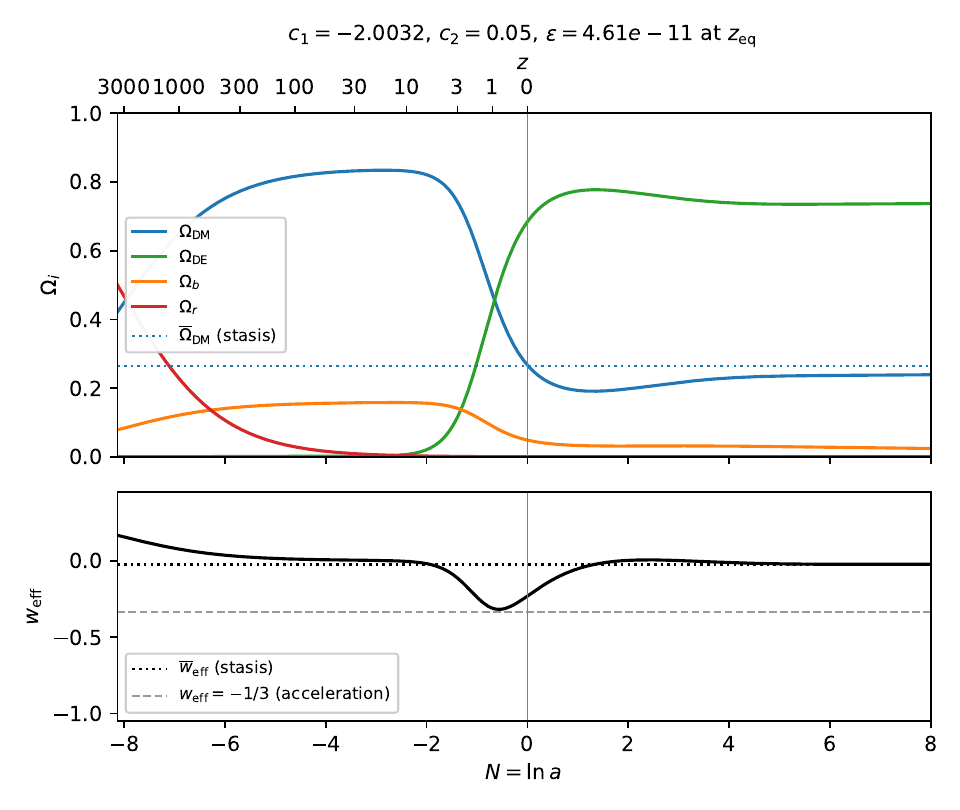}
    \caption{Anchored evolution of the abundances (top panels) and of the total
    equation of state (bottom panels) from matter-radiation equality to eight e-folds
    beyond the present, for the benchmark points $\swDM = 0.264$, $c_2 = 0.05$ with
    $c_1 = 2.035$ (top) and $c_1 = -2.003$ (bottom).}
    \label{fig:evolution}
\end{figure}

\Cref{fig:evolution} shows the anchored evolution for two benchmark points with $\swDM = 0.264$, one on each branch $c_1 > 0$ and $c_1 < 0$. The abundances reproduce the observed history up to the present, and the trajectories are then captured into stasis a few e-folds later. However, the total equation of state never reaches the observed value $\weff = -\wDE^{(0)} \simeq -0.68$ today.

\begin{figure}[htbp]
    \centering
    \includegraphics[width=0.8\linewidth]{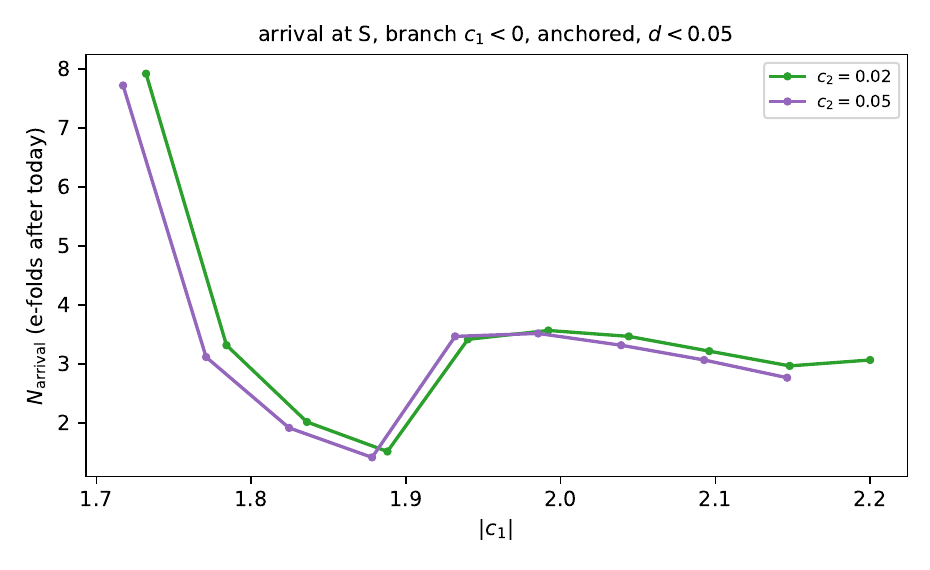}
    \caption{The number of e-folds after today at which the anchored trajectory comes
    within $0.05$ of the stasis point S in the $(u, \wDM)$ plane, on the branch
    $c_1 < 0$, for two values of the coupling. The approach is a spiral, so the curve
    is not monotonic in $c_1$, and it rises toward the stasis boundary where the
    convergence rate vanishes. The range ends where the anchoring fails because the
    trajectory is captured before $\wDE$ reaches its observed value.}
    \label{fig:arrival}
\end{figure}

On the branch $c_1 < 0$ the point S is an attractor, so once stasis is reached it lasts forever, and the relevant question is how soon it is reached. For each $(c_1, c_2)$ inside the stasis region we integrate the anchored solution forward and record the number of e-folds after the present at which the trajectory comes within a distance $0.05$ of S in the $(u, \wDM)$ plane. The result is shown in \cref{fig:arrival}. Across the region the arrival takes between $1$ and $4$ e-folds, and near the boundary $c_1(c_1 - c_2) = 3$ where the convergence rate $\gamma$ of \cref{eq:gamma_def} vanishes it takes $7$ to $8$ e-folds. If the parameters of the model lay in this region, the universe would be captured into eternal stasis in the near future. Whether this range of parameters is consistent with current observational constraints is the focus of the next subsection.

\subsection{An acceleration bound on the stasis region}\label{sec:accel}
\begin{figure}[htbp]
    \centering
    \includegraphics[width=0.85\linewidth]{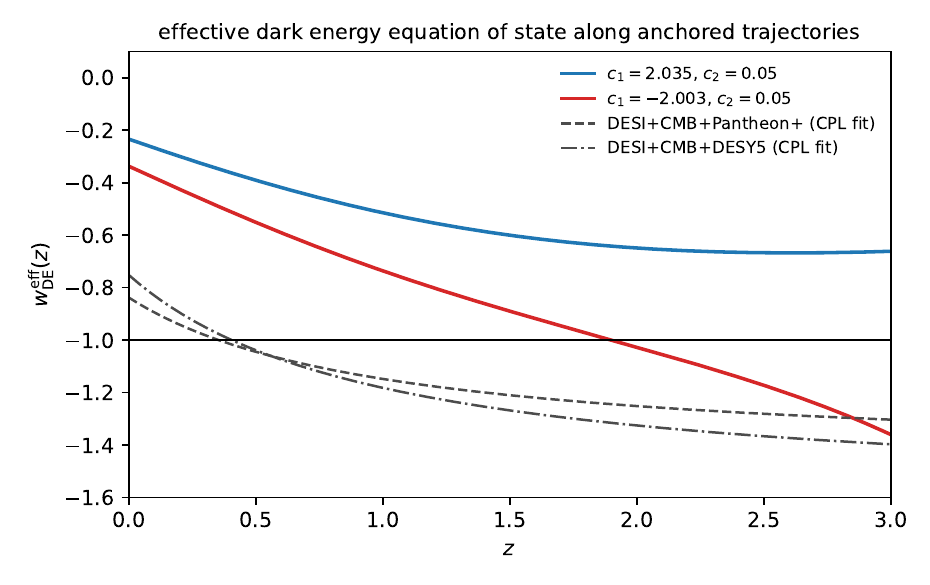}
    \caption{The effective dark energy equation of state \cref{eq:eff_split} along
    the two anchored benchmark trajectories of \cref{fig:evolution}. The dashed and dash dotted
    curves are the $w_0 w_a$ best fits of \cite{DESI:2025zgx} for DESI combined with the CMB
    and with the Pantheon+ and DESY5 supernova samples.}
    \label{fig:wde_eff}
\end{figure}

In the anchored cosmology, the total equation of state today follows from \cref{eq:weff_state} and the definition \cref{eq:DE_def} of $\wDE$,
\begin{equation}
    \weff(0) = \frac{u_0^2}{3} - \wDE^{(0)} + \frac{\wR^{(0)}}{3} \simeq \frac{u_0^2}{3} - \wDE^{(0)}\;,
\end{equation}
where $u_0$ is the field velocity at the time when $\wDE$ reaches its measured value today and the radiation term, $\wR^{(0)} \simeq 9\times 10^{-5}$, is dropped. Note that $\wDE^{(0)}$ is the full radion abundance including its kinetic energy, which is the quantity fixed by the anchoring. From the above expression for $\weff(0)$, we have the bound
\begin{equation}\label{eq:floor}
    \weff(0) \geq -\wDE^{(0)} = -0.685\;,
\end{equation}
which is saturated only by a field which is still frozen, $u_0 = 0$, for which the dark energy is pure potential energy. More generally, \cref{eq:weff_state} gives $\weff = u^2/6 - \wV + \wR/3 \geq -\wV \geq -1$ at every time, with equality only in exact de Sitter space, and the radion's own equation of state $w_\phi = (u^2/6 - \wV)/(u^2/6 + \wV)$ obeys the same bound. This is the null energy condition for a canonical scalar with positive potential, so nothing in the physical description of this model is ever phantom. However, the effective dark energy of \cref{eq:eff_split} can be as shown in \cref{fig:wde_eff}. On the branch $c_1 < 0$ the field rolls toward smaller $\phi$, the tower mass grows, and the physical dark matter diluted more slowly than $a^{-3}$ in the past. As a result, $\rho_{\m{DE}}^{\m{eff}}$ was smaller than $\rho_\phi$ and $w_{\m{DE}}^{\m{eff}}$ crosses $-1$ at $z \simeq 1$ to $2$, while on the branch $c_1 > 0$ the opposite happens and $w_{\m{DE}}^{\m{eff}} > -1$ throughout. This is how coupled dark matter and dark energy exhibit apparent phantom behavior as noted in \cite{Bedroya2025,Agrawal:2019dlm}, see also \cite{Das:2005yj,GomezValent:2026dea}.

\begin{figure}[htbp]
    \centering
    \includegraphics[width=0.85\linewidth]{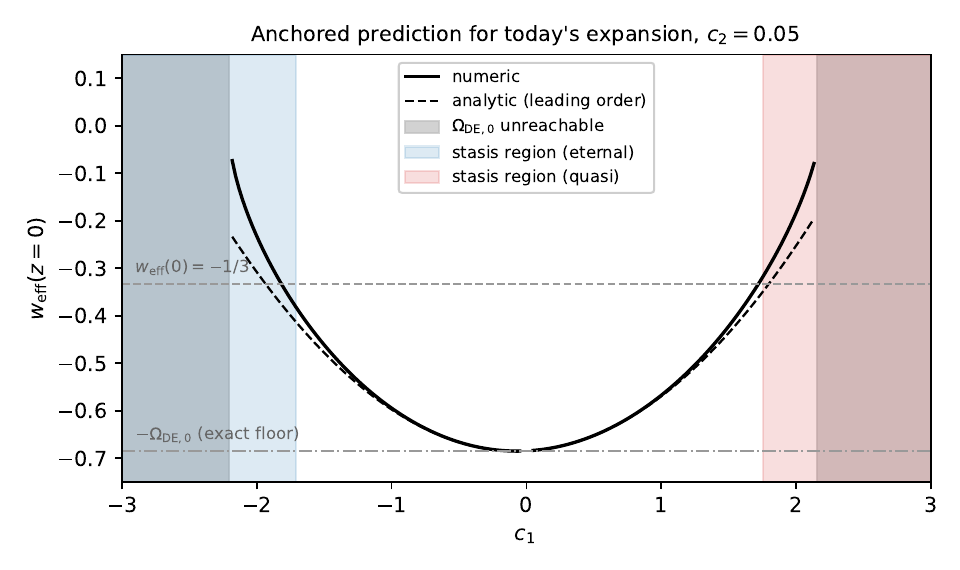}
    \caption{The anchored prediction for the total equation of state today as a
    function of $c_1$ at $c_2 = 0.05$ (solid), together with the analytic result
    \cref{eq:thawing_u} (dashed), the acceleration threshold $-1/3$, the bound
    \cref{eq:floor}, and the stasis regions of both branches (shaded). In the grey
    regions the value $\wDE^{(0)} = 0.685$ cannot be reached for any initial
    condition.}
    \label{fig:weff0_slice}
\end{figure}

To study what conditions are required to find an accelerating universe today, we linearize the anchored evolution about the frozen field, as described in \cref{app:thawing}. This generalizes the thawing expansion of \cite{Scherrer:2007pu,Dutta:2008qn,Chiba:2009sj} to the coupled model. We find
\begin{equation}\label{eq:thawing_u}
    u(\wDE) = c_1\,\big[1 - T(\wDE)\big] + 2c_2\big[1 - F_{\m{b}}\big]\,T(\wDE)\;,
\end{equation}
where $F_{\m{b}} = 0.157$ is the baryon fraction of matter, and
\begin{equation}
    T(\wDE) = \frac{(1-\wDE)\,\m{arctanh}\sqrt{\wDE}}{\sqrt{\wDE}}\;.
\end{equation}
Only the coupled fraction of the dust drives the field, which is the origin of the factor $1 - F_{\m{b}}$. At $\wDE = 0.685$ one has $T = 0.449$, and the requirement of acceleration today, $\weff(0) < -1/3$, becomes
\begin{equation}\label{eq:accel_bound}
    \big|\,c_1\,(1 - T_0) + 2c_2(1-F_{\m{b}})\,T_0\,\big| < \sqrt{3\big(\wDE^{(0)} - \tfrac{1}{3}\big)}\;,
\end{equation}
which reads
\begin{equation}
    \big|c_1 + 1.38\,c_2\big| \lesssim 1.87\;.
\end{equation}
We have verified \cref{eq:thawing_u} against full integrations. It is accurate to better than a percent for $|c_1| \lesssim 1.2$ and to about $10$ percent at the stasis boundary, see \cref{fig:weff0_slice}.

\begin{figure}[htbp]
    \centering
    \includegraphics[width=0.49\linewidth]{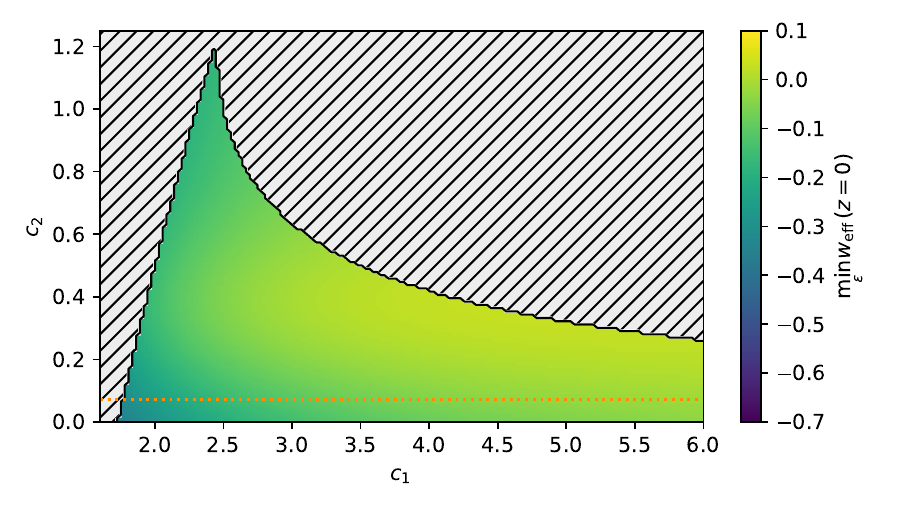}\hfill
    \includegraphics[width=0.49\linewidth]{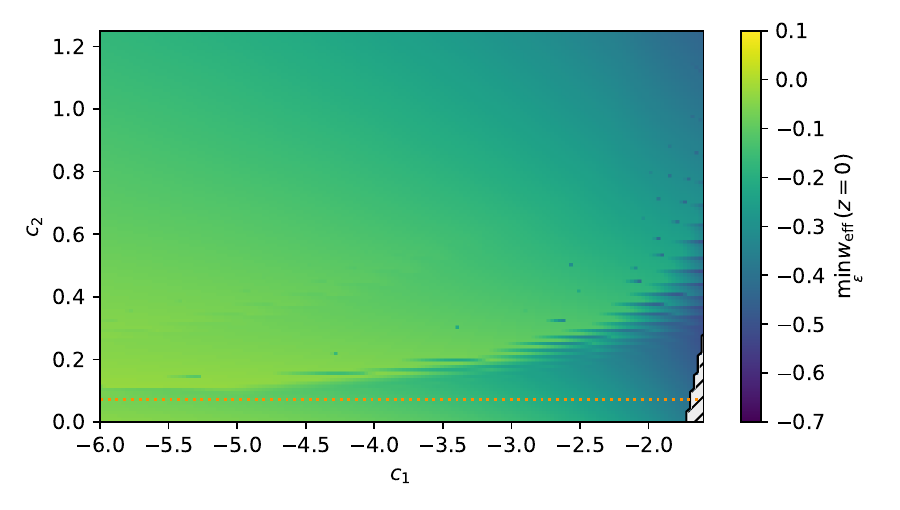}
    \caption{The most negative total equation of state which can be reached today
    over all initial conditions, within the stasis regions of the two branches, for
    $c_1 > 0$ (left) and $c_1 < 0$ (right). On the branch $c_1 > 0$ the value never
    falls below the acceleration threshold $-1/3$ at couplings allowed by the fifth-force
    bound (orange dotted line), and on the branch $c_1 < 0$ it does so only in the thin
    strip at the boundary discussed in the text, never reaching the observed $-0.685$.
    The striping seen on the branch $c_1 < 0$ arises because trajectories there spiral into S before
    today, so that $\weff(0)$ oscillates with the initial condition and the minimum
    over a discrete set of initial conditions is aliased.}
    \label{fig:weff0min}
\end{figure}

Now compare this bounds with the stasis region for $V_0 > 0$ defined by \cref{eq:stasis_V0_pos}. Stasis requires $|c_1| \gtrsim \sqrt{3} \approx 1.73$ at small $c_2$, while acceleration requires $|c_1| \lesssim 1.7$ to $1.9$. Numerically, at $c_2 = 0.05$ the anchored universe accelerates today only for
\begin{equation}
    -1.82 < c_1 < 1.72\;,
\end{equation}
while stasis exists only for $c_1 \geq 1.76$ or $c_1 \leq -1.71$. The stasis region on the branch $c_1 > 0$ never accelerates at any epoch. The region on the branch $c_1 < 0$ survives only in a thin strip $-1.82 \leq c_1 \leq -1.71$ at the boundary, with $\weff(0) \approx -0.36$, which is half of the observed value, and with $\swDM \lesssim 0.11$. For any $|c_1| \gtrsim 2.2$ the anchoring fails at the first step, since the trajectory is captured into stasis with $\swDE < 0.685$ before the dark energy reaches its observed abundance. Those parameters cannot produce the dark energy budget observed today for any initial condition. \Cref{fig:weff0min} shows the acceleration constraint by itself over both branches, minimized over all initial conditions.

Combining \cref{eq:fifthforce} with \cref{eq:accel_bound} and with the requirement that the dark energy abundance can be reached, we find that the stasis region of the coupled dark matter and dark energy model is incompatible with the observed accelerating universe, independent of initial conditions. The only exception is the thin strip at the boundary of the branch $c_1 < 0$, which is itself disfavored by the measured value of $\weff(0)$.

\subsection{Capture into stasis in the future}\label{sec:capture}
The exclusions above assume constant slopes. By the integrability condition \cref{eq:integrability}, running slopes leave one question open. Even if the slopes are shallow today, as the data require, the field rolls toward $\phi \rightarrow \infty$. If $c_1(\phi)$ becomes steep enough, the trajectory enters the stasis region and is captured. Acceleration then ends and the universe settles into stasis. Whether this happens depends on the asymptotic form of the potential.

At small coupling $c_2$ the criterion for capture is
\begin{equation}
    c_{1,\infty} \gtrsim \sqrt{3}\;,
\end{equation}
which is the threshold for the existence of stasis. Potentials with asymptotic slope $\sqrt{2}$, such as the potentials of sech type found consistent with DESI in \cite{Bedroya2025}, end acceleration but are not captured. We return to the significance of this value in \cref{sec:swamp}.

\begin{figure}[htbp]
    \centering
    \includegraphics[width=0.8\linewidth]{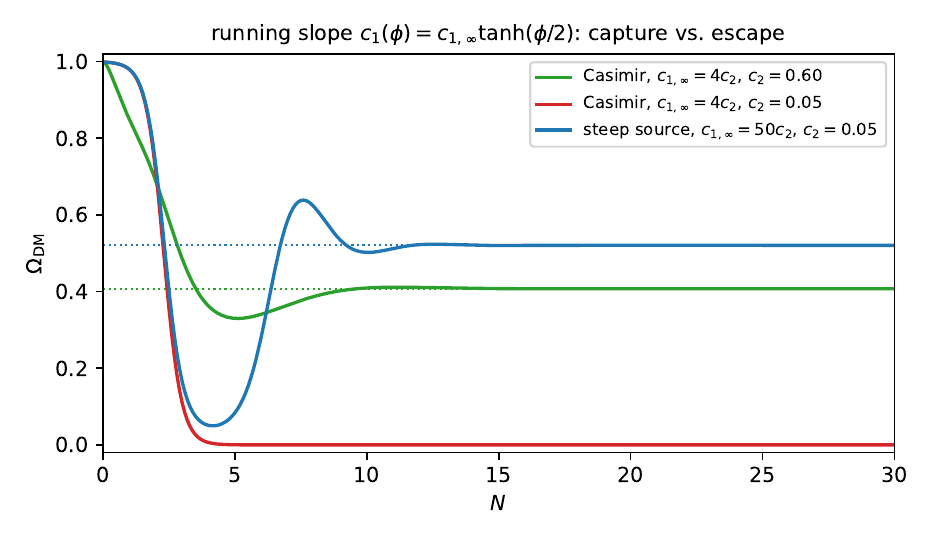}
    \caption{Capture into stasis in the future for the running slope
    $c_1(\phi) = c_{1,\infty}\tanh(\phi/2)$. Green shows a Casimir asymptote $c_{1,\infty} = 4c_2$ at
    $c_2 = 0.60$, which is captured into the stasis with $\sweff = 1/3$ (dotted
    target). Red shows the same asymptote at the allowed coupling
    $c_2 = 0.05$, for which the dark matter fades and the universe accelerates
    forever. Blue shows a hypothetical steep source with $c_{1,\infty}/c_2 = 50$, which is
    captured at small coupling (dotted target).}
    \label{fig:capture_demo}
\end{figure}

Consider sources that are powers of the radius $R$,
\begin{equation}
    V \propto R^{-p}\;,
\end{equation}
which give straight lines $c_1 = p\,c_2$ in the plane of slopes since $M \propto 1/R$ ties both slopes to $\d\ln R/\d\phi$. Capture which is consistent with the fifth-force bound \cref{eq:fifthforce} then requires
\begin{equation}\label{eq:capture_nogo}
    \frac{c_{1,\infty}}{c_2} \gtrsim \frac{\sqrt{3}}{0.071} \approx 24\;.
\end{equation}
The physically motivated sources have $p = 4$ for the Casimir energy \cite{Arkani-Hamed:2007ryu,Montero:2022prj}, $p = 2$ for internal curvature, and single digit values for fluxes, $p = (2q + n)/(1 + n/2)$ for a $q$-form flux on $n$ internal dimensions. The latter two follow from the Einstein frame scalings of \cite{Hertzberg:2007wc,Danielsson:2009ff} with the string coupling held fixed, which give $V \propto m_{\m{KK}}^2$ for curvature and $p = 3/2$ to $9/2$ for the fluxes of type IIA on a six dimensional internal space. See also \cite{Blumenhagen:2022zzw} for an example of scalar potentials generated in stringy constructions motivated by the DD scenario. In particular, for the Casimir potential of the DD \cite{Montero:2022prj} one has $c_1(\phi) \rightarrow 4c_2$ asymptotically, and the window of couplings in which capture occurs is
\begin{equation}
    \frac{1}{2} < c_2 < \frac{1}{\sqrt{2}}\;,
\end{equation}
which is excluded today by \cref{eq:fifthforce}. We note in passing that $c_1 = 4c_2$ is also the line on which RS and KRD exchange stability, see \cref{tab:stability}, and on which the stasis S has $\sweff = 1/3$. The pure unwarped radion with $c_2 = \sqrt{3/2}$ is not captured either. \Cref{fig:capture_demo} shows the three behaviors with explicit running slope trajectories. On the line $c_1 = 4c_2$ of a pure Casimir potential, at $c_2 = 0.6$, the trajectory is captured into a stasis with $\sweff = 1/3$, so that the universe expands forever as if it were radiation dominated while it is made of dark matter and dark energy. At the allowed coupling $c_2 = 0.05$ the trajectory escapes to eternal acceleration. For a hypothetical steep source with $c_{1,\infty}/c_2 = 50$ the trajectory is captured at small coupling.

There are two ways in which this conclusion could fail, a potential with several partially canceling sources whose local slope becomes steep over a limited range of the field, as in the asymptotic flux potentials of \cite{Calderon-Infante:2022nxb}, and a coupling $c_2(\phi)$ which grows asymptotically into the window above. We return to both in \cref{sec:conc}. Either can be checked with the formalism of \cref{sec:theorem} for any concrete construction.

\section{Stasis with dark matter production}\label{sec:production}

In the DD, energy can also be pumped into the DM sector by the cooling of the SM brane. The tower of dark gravitons is populated by this cooling with a production rate \cite{Gonzalo:2022jac}
\begin{equation}\label{eq:prod_rate}
    \left.\frac{\d\rho_{\m{KK}}}{\d t}\right|_{\m{prod}} \sim \frac{T^{8}}{\hat{M}_p^{3}} \sim \frac{T^8}{M(\phi)}
\end{equation}
for one extra dimension, where $\hat{M}_p^3 = M_{\m{pl}}^2/(2\pi R) \sim M$ in our units $M_{\m{pl}} = 1$. The energy which enters the tower is drawn from the SM radiation, so the source in \cref{eq:aut_dm} is accompanied by a sink in \cref{eq:aut_r}. The system remains autonomous if we introduce one new variable,
\begin{equation}
    \chi \equiv \frac{H}{M}\;,\qquad
    \chi' = \chi\,\Big[\,c_2u - \tfrac{3}{2}(1+\weff)\,\Big]\;,
\end{equation}
see \cref{app:aut_ODEs} for more details. The production per e-fold is
\begin{equation}
    s \equiv \frac{S}{3H^3} = 3\kappa\,\wR^2\,\chi\;,
\end{equation}
where $\kappa$ collects the degrees of freedom on the brane. The extended system is written out in \cref{app:aut_ODEs}. It has new fixed points with $s \neq 0$, which we will refer to as ``production stasis'' fixed points. The conditions for these fixed points are
\begin{equation}\label{eq:prod_pins}
    c_2u = \frac{3}{2}\big(1 + \sweff\big)\;,\qquad
    s = \wR\big(3\sweff - 1\big)\;,\qquad \sweff \in \big(\tfrac{1}{3}, 1\big)\;,
\end{equation}
together with $c_1 = 2c_2$ if the potential participates.

At a fixed point the abundances are constant, so by the radiation equation the fixed point requires the production per e-fold $s = S/3H^3$ to be constant and the source must then scale as $H^3$. In a radiation dominated epoch with $H^2 \sim \rho_{\m{r}} \sim T^4$, this means $S \sim T^6$. The actual source described above scales as $T^8/M$, so at fixed mass $s \sim T^2/M \sim H/M$ and the source dies away as the universe expands. Production stasis in the sense described above is therefore impossible in models with a fixed dark matter mass. The first condition in \cref{eq:prod_pins} states that the tower mass decreases in step with the Hubble rate, so that $H/M$ stays constant. The radion is exactly the ingredient which makes a balance between production and dilution possible.

However, the DD scenario as typically realized is far from this balance. Take the SM radiation to dominate the expansion during the cooling, as in \cite{Gonzalo:2022jac}. Then, $H \sim T^2$ with a brane temperature of a few GeV and $M \sim 0.1\,$eV, the present value of the KK scale which the frozen radion preserves back to the epoch of production. We then find
\begin{equation}
    \frac{H}{M} \sim \frac{T^2}{M} \sim 10^{-8} \text{ to } 10^{-7}
\end{equation}
at the epoch of production, so $s \ll 1$ at all times. The production is the UV dominated burst described in \cite{Gonzalo:2022jac}, and the relic abundance is set by the integral of the burst rather than by an attractor. The balance becomes relevant only when $s$ is of order one, which for $M \sim 0.1\,$eV requires brane temperatures of order $\sqrt{M M_{\m{pl}}/\kappa} \sim$ TeV, well above the initial temperatures of the scenario. In that regime the burst picture no longer applies since the sink in the radiation equation is no longer negligible. Whether the trajectory is captured by the production stasis, which would cap the dark matter abundance, depends on its stability which we have not analyzed. Such temperatures lie outside the scenario as formulated. The initial temperature is fixed near a few GeV both by the relic abundance and by the requirement that the moduli have settled \cite{Gonzalo:2022jac}, and the overproduction of bulk gravitons bounds the brane temperature in extra dimensional models in general \cite{Arkani-Hamed:1998sfv,Hall:1999mk}. However, this regime is not without interest. It has been pointed out that a brane temperature in the GeV range or above is compatible with these bounds if the gravitons produced at that temperature decay to the SM before recombination \cite{Macesanu:2004gf}. Such decays of a tower into radiation are the energy transfer of the original stasis mechanism \cite{Dienes:2021woi}, so a hot brane would combine the production considered here with a decay channel and the two together could support stasis solutions beyond those of \cref{eq:prod_pins}. We leave the study of such regimes in which stasis before BBN with $\weff$ between $1/3$ and $1$ arises and exits by itself when the source dies to future work.

\section{Relation to swampland bounds}\label{sec:swamp}

Whether string theory admits de Sitter vacua is an actively debated question. Constructions such as \cite{Kachru:2003aw} argue that it does, while others argue that de Sitter space is difficult to obtain in a controlled way and that quintessence is the natural alternative \cite{Danielsson:2018ztv,Cicoli:2018kdo,Agrawal:2018own}. The de Sitter conjecture and the trans-Planckian censorship conjecture (TCC) bound the asymptotic slope of the potential \cite{Obied:2018sgi,Ooguri:2018wrx,Garg:2018reu,Bedroya:2019snp,Rudelius:2022gbz,Etheredge:2026cha}. In $d = 4$ these give
\begin{equation}
    \lambda \geq \sqrt{2}
\end{equation}
in infinite distance limits and the weaker bound
\begin{equation}
    \lambda_{\m{interior}} \geq \sqrt{\frac{2}{3}}
\end{equation}
in the interior of moduli space. The Sharpened Distance Conjecture (SDC) bounds the decay rate of towers \cite{Ooguri:2006in,Etheredge:2022opl}, which in $d = 4$ reads
\begin{equation}
    \frac{1}{\sqrt{2}} \leq \lambda' \leq \sqrt{\frac{3}{2}}\;.
\end{equation}
We identify $\lambda = c_1$ and $\lambda' = c_2$ for small excursions in field space. The structure of the stasis solution space lines up with these bounds in several ways.
\begin{itemize}
    \item On the branch $c_1 > 0$ the maximal $c_2$ which admits stasis is exactly the SDC upper bound. The apex of the stasis region is at $c_2 = \sqrt{3/2}$ and $c_1 = 2\sqrt{3/2}$, where $\swDM \rightarrow 0$.
    \item The threshold for the existence of the radiation era stasis KRD is exactly the SDC lower bound $c_2 = 1/\sqrt{2}$. The window for capture of a pure Casimir potential ends at the same value.
    \item The minimal asymptotic slope $\sqrt{2}$ of the TCC is below the capture threshold $\sqrt{3}$, so potentials which only saturate the TCC end acceleration without stasis.
\end{itemize}
We note that sharp numerical SDC bounds are not firmly established in the presence of scalar potentials and warping, so these identifications should be read with this caveat. Since cosmological data does not show $|\nabla V|/V \geq \sqrt{2}$, our universe is not currently in an asymptotic regime of scalar field space and only the interior bound applies today. The preferred value $|c_1| \sim 0.8$ of \cite{Bedroya2025} lies close to this interior bound.

\section{Conclusions and future work}\label{sec:conc}

In this work we have analyzed cosmological stasis in the coupled dark matter and dark energy cosmology of the DD scenario, keeping baryons and radiation throughout. The known results we build on are the stationary solutions of coupled quintessence \cite{Wetterich:1994bg,Amendola:1999er,Amendola:2000ub,Amendola:2001rc} and the fact that a single coupled fluid requires exponential $V$ and $M$ for such solutions \cite{Liddle:1998xm,Amendola:2006qi}. The new results of this work are the following.
\begin{itemize}
    \item We define exact stasis as a fixed point of the full system with spectators. We show that for stasis solutions with arbitrary $V$ and $M$, the slopes must be constant on the segment of field space which is traversed. When the slopes run, the outcome is either exact stasis or a drifting quasi-stasis depending on whether the running is integrable. Sums of exponentials converge to stasis solutions while potentials with polynomial prefactors do not.
    \item The spectators determine the lifetime and the exit of every stasis solution. We obtain a closed form lifetime for the coupled stasis S and validate it numerically. With the baryon fraction of our universe there is no stasis or quasi-stasis achieved on the branch $c_1 > 0$.
    \item Anchoring the cosmology to the measured abundances at matter-radiation equality and today removes all freedom for fixed slopes $c_1$ and $c_2$. This gives the exact bound $\weff(0) \geq -\wDE^{(0)}$ and a closed form for the field velocity today which generalizes the thawing expansion to the coupled model. The stasis region cannot accelerate.
    \item Capture into stasis in the future is decided by the asymptotic slope ratio $c_{1,\infty}/c_2$. No geometric source of the radion potential satisfies the requirement once the fifth-force bound is imposed.
    \item The production of dark matter from the brane closes into an autonomous system with one extra variable. It supports a stasis solution sustained by the mass drift which the typical DD scenario does not reach.
\end{itemize}
The verdict for whether stasis within the considered model is viable for our universe is negative in each epoch, and for a different reason in each. We summarize the reasons and then ask what could change them.

In the early universe, the stasis solutions which contain radiation are RS and KRD. Both hold a large fraction of the energy density outside the SM and are excluded by BBN if they persist to that epoch, see \cref{sec:early}. The deeper problem is that in this model neither has an exit into the standard cosmology. In the stasis models of \cite{Dienes:2021woi,Dienes:2023ubz} the epoch ends when the tower which sustains it is exhausted, whereas here the drift of the tower mass never switches off, and the only exits are the growth of baryons toward RS or toward a baryon dominated universe without dark matter. Standard early cosmology therefore requires the radion to be frozen through the radiation era.

In the present epoch the obstacle is acceleration. Stasis exists only for $|c_1| \gtrsim \sqrt{3}$ at the couplings allowed by fifth-force bounds, while the anchored cosmology accelerates today only for $|c_1 + 1.38c_2| \lesssim 1.87$. The exact bound $\weff(0) \geq -\wDE^{(0)}$ shows that no choice of initial conditions can help. The two regions overlap only in a thin strip at the boundary of the branch $c_1 < 0$ where $\weff(0) \approx -0.36$, and on the branch $c_1 > 0$ the baryon fraction of our universe prevents even a quasi-stasis. The total equation of state is never phantom, while the effective dark energy of the $\Lambda$CDM split crosses $-1$ in the past on the branch $c_1 < 0$, which is the origin of the apparent phantom behavior of the scenario. In that split the stasis region has $w_{\m{DE}}^{\m{eff}}(0) \gtrsim -0.56$, far from the DESI preferred values. The universe is therefore not in stasis today, and the coincidence problem cannot be attributed to a stasis attractor in this model.

In the future the question is whether the radion is captured into the eternal stasis of the branch $c_1 < 0$ as it rolls. For the geometric potentials, capture is decided by the asymptotic slope and the requirement $c_{1,\infty}/c_2 \gtrsim 24$ of \cref{sec:capture} cannot be met by Casimir, curvature, or flux contributions once the fifth-force bound is imposed. The universe therefore ends in eternal acceleration rather than in stasis. The closest stasis to the observed universe is the boundary of the eternal region at $c_1 \approx -1.7$ and $c_2 \approx 0.05$, a few standard deviations beyond the posteriors of \cite{Bedroya2025}. This is a falsifiable target as supernova and BAO data improve.

Each of these obstacles is tied to a specific simplification of the model, and the DD contains ingredients which could overcome each of them. The model studied here contains a single KK tower and a single transfer mechanism, the drift of its mass. The DD scenario contains more. With a gauged $B{-}L$ symmetry in the bulk \cite{Montero:2025hye} there are KK towers of right handed neutrinos with masses in the keV range and a bulk $B{-}L$ gauge boson, both of which decay to SM states. A tower of bulk states decaying to radiation is precisely the ingredient of the original stasis mechanism \cite{Dienes:2021woi}, and it supplies what the radion cannot, namely a transfer from matter to radiation which ends by itself when the tower is exhausted. Combined with the mass drift of the radion this would give a two-source stasis in the early universe with a natural exit, and since the decaying towers populate the radiation bath rather than the dark sector, the BBN obstacle to RS and KRD would be modified as well. Whether the exit occurs early enough is a question about the tower spectrum and lifetimes, which are fixed in the $B{-}L$ construction. An early stasis of this type would hold a fraction of the energy density in the dark sector during the radiation era and release it before recombination, which is the behavior of early dark energy \cite{Poulin:2018cxd,Poulin:2023lkg}. Whether a stasis epoch can reproduce the fraction and the timing that relieve the Hubble tension is a concrete question which the formalism of this paper can address.

For the present epoch the obstacle is that the coupled dark matter is all of the dark matter, so that the fifth-force bound $c_2 \lesssim 0.071$ applies to the full abundance and the accelerated stationary attractor of \cite{Amendola:2001rc} at strong coupling is excluded. If the KK modes are only a fraction of the dark matter with the remainder in axions (see \cite{Gendler:2024gdo,Langhoff:2026qqg,Li:2024jko} for various works studying axions in the DD), sterile neutrinos, or primordial black holes, the fifth-force bound applies to the coupled fraction alone and weakens accordingly \cite{Archidiacono:2022iuu}. A subdominant coupled component with a larger $c_2$ reopens the strongly coupled region, where stasis can accelerate. This further changes the anchoring, since only part of the dark matter directly drives the field $\phi$. The acceleration bound of \cref{sec:accel} would have to be rederived with a mixed dark matter sector, and we regard this as the most promising route to a viable late time stasis in the scenario.

For the future, the loopholes identified in \cref{sec:capture} remain. Potentials with several sources which partially cancel can become steep over a limited range of the field, turning eternal capture into a stasis epoch of finite duration with entry and exit generated by the potential itself. A coupling $c_2(\phi)$ which grows asymptotically would reopen capture for a pure Casimir potential. Both can be checked for any concrete construction by computing the running slopes and propagating them through the drift formalism of \cref{sec:theorem}.

Finally, the production of dark matter from the brane supports a production stasis which the scenario does not reach. The balance requires brane temperatures of order $10\,$TeV while the standard DD scenario fixes the initial temperature near a few GeV. A hotter brane is compatible with graviton overproduction bounds only if the produced gravitons decay to the SM before recombination. Such decays of a tower into radiation are the transfer of the original stasis mechanism, so production and decay would act together. A stasis of this kind before BBN would end by itself when the source dies. It is a second candidate for an early epoch of the kind discussed above.

\section*{Acknowledgments}
AS would like to thank Alessandro Borys, Victoria Knapp-P\'erez, Arvind Rajaraman, Sanjay Raman, Michael Ratz, and Tim M.P. Tait for helpful comments regarding this work. AS would also like to thank the organizers of the 23rd Simons Physics Summer Workshop, where part of this work was conducted. AS acknowledges the use of Anthropic's Claude to generate numerical code for creating plots, and to find grammar and spelling mistakes in earlier drafts of this work. This material is based upon work supported by the National Science Foundation Graduate Research Fellowship Program under Grant No. DGE-2235784. Any opinions, findings, and conclusions or recommendations expressed in this material are those of the authors and do not necessarily reflect the views of the National Science Foundation.

\begin{appendix}

\section{Friedmann and fluid equations}\label{app:FEqns}
We work in units $M_{\m{pl}}^2 = (8\pi G)^{-1} = 1$. The Friedmann equations are derived from the field equations corresponding to the FLRW metric with the stress-energy tensor for a perfect fluid. The metric is taken to be (setting $c=1$)
\begin{equation}
	-\dd s^2 = a(t)^2 \dd{s}_3^2 - \dd{t}^2\;,
\end{equation}
with $a(t)$ the scale factor and $\dd{s}_3^2$ a three-dimensional metric corresponding to flat space, spherical space with constant positive curvature, or hyperbolic space with constant negative curvature. The $00$ component of the field equations yields
\begin{equation}\label{eq:FirstFE}
	H^2 = \frac{1}{3}\rho - \frac{k}{a^2}\;,
\end{equation}
where $\rho$ is the energy density of the perfect fluid, $k$ takes on the values $0$, $1$, and $-1$ for flat, spherical, or hyperbolic spaces respectively, and $H \equiv \dot{a}/a$ is the Hubble parameter. In the main text, we consider a flat FLRW universe, so $k=0$ and the curvature term vanishes. Combining the above result with the trace of the field equations yields one more independent equation
\begin{equation}\label{eq:SecondFE}
	\dot{H} + H^2 = -\frac{1}{6} (  \rho + 3p )\;,
\end{equation}
where $p$ is the pressure of the perfect fluid. We will refer to \cref{eq:FirstFE} as the Friedmann equation and \cref{eq:SecondFE} as the Friedmann acceleration equation. The pressure and density of a perfect fluid are related through the equation of state
\begin{equation}
	w = \frac{p}{\rho}\;,
\end{equation}
with $w=1/3$ for radiation and $w=0$ for matter. For vacuum energy, $w = -1$.

The fluid equation will follow from taking the divergence of the field equations and applying Bianchi identities to conclude
\begin{equation}
	\nabla_\mu T^{\mu\nu} = 0\;,
\end{equation}
which is just a statement of the conservation of energy and momentum in curved spacetime. Tracing over this result (assuming an FLRW solution for the metric and assuming that $T^{\mu\nu}$ is the stress energy of a perfect fluid with equation of state $w$), one finds
\begin{equation}
	\dot{\rho}_i = -3H(1+w)\rho_i \;.
\end{equation}
When there are other contributions to the energy densities (such as the decay of massive particles into radiation, which removes energy from $\rho_M$ and adds to $\rho_\gamma$), the fluid equation will need to be modified to ensure energy conservation.

Combining \cref{eq:FirstFE,eq:SecondFE} in the flat case gives
\begin{equation}
    \dot{H} = -\frac{1}{2}\big(\rho + p\big)\;,
\end{equation}
and hence the identity
\begin{equation}\label{eq:HpH_weff}
    \frac{H'}{H} = \frac{\dot{H}}{H^2} = -\frac{3}{2}\big(1 + \weff\big)\;,
\end{equation}
which is used throughout the main text.

\section{Derivation of the autonomous system}\label{app:aut_ODEs}
To convert from $t$ to $N = \log a$, we note that
\begin{equation}
    \frac{\d}{\d t} = H\frac{\d}{\d N}\;,\qquad
    \frac{\d^2}{\d t^2} = H^2\bigg(\frac{\d^2}{\d N^2} + \frac{H'}{H}\frac{\d}{\d N}\bigg)\;.
\end{equation}
The scalar equation of motion \cref{eq:phi_eom} becomes
\begin{equation}\label{eq:scalar_eom_N}
    \phi'' + \phi'\Big(3 + \frac{H'}{H}\Big) + \frac{1}{H^2}\frac{\d V}{\d\phi} + \frac{\pDM}{H^2}\,\frac{\d \log M}{\d\phi} = 0\;.
\end{equation}
By \cref{eq:HpH_weff} the friction coefficient is
\begin{equation}
    3 + \frac{H'}{H} = \frac{3}{2}\big(1 - \weff\big)\;.
\end{equation}
The remaining terms are
\begin{equation}
    \frac{1}{H^2}\frac{\d V}{\d\phi} = -3c_1\,\wV\;,\qquad
    \frac{\pDM}{H^2}\,\frac{\d \log M}{\d\phi} = -3c_2\wDM\;.
\end{equation}
Setting $u = \phi'$ gives \cref{eq:aut_u}.

For the abundances, differentiating $\Omega_i = \rho_i/3H^2$ with respect to $N$ and using \cref{eq:HpH_weff} gives
\begin{equation}\label{eq:Omega_generic}
    \Omega_i' = \Omega_i\Big(\frac{\rho_i'}{\rho_i} + 3(1 + \weff)\Big)
\end{equation}
for any component. For the baryons and radiation, $\rho_i'/\rho_i = \dot{\rho}_i/H\rho_i$ is $-3$ and $-4$ respectively. For the dark matter, \cref{eq:DMfluid} gives $\rho_{\m{DM}}'/\pDM = -3 - c_2u$. Inserting these into \cref{eq:Omega_generic} gives
\begin{equation}
    \wDM' = \wDM\big(3\weff - c_2u\big)\;,\qquad
    \wB' = 3\weff\,\wB\;,\qquad
    \wR' = \wR\big(3\weff - 1\big)\;,
\end{equation}
which are \cref{eq:aut_dm,eq:aut_b,eq:aut_r}. Together with \cref{eq:aut_u} these four equations close, since $\weff$ is a function of $(u, \wDM, \wB, \wR)$ by \cref{eq:weff_state}. The potential fraction $\wV$ is fixed by the constraint and its evolution is a consequence of the system. It is nevertheless useful to have it in the same form, since $V$ depends on $N$ only through $\phi$ and therefore $\rho_V'/\rho_V = -c_1u$, so that \cref{eq:Omega_generic} gives \cref{eq:aut_V}. One can check that \cref{eq:aut_V} agrees with differentiating $\wV = 1 - u^2/6 - \wDM - \wB - \wR$ along the flow, which is the statement that the flow preserves the constraint \cref{eq:constraint}.

In terms of the state variables,
\begin{equation}
    \frac{H'}{H} = -\frac{1}{2}u^2 - \frac{3}{2}\big(\wDM + \wB\big) - 2\wR\;,
\end{equation}
which agrees with \cref{eq:HpH_weff} and \cref{eq:weff_state}. Each face of the boundary of the state space is invariant, since every equation above is proportional to its own abundance. On the face $\wB = \wR = 0$ the system reduces to
\begin{subequations}\label{eq:aut_sys2}
\begin{align}
    u' &= -3u\Big(1 - \frac{u^2}{6} - \frac{\wDM}{2}\Big) + 3c_1\Big(1 - \frac{u^2}{6} - \wDM\Big) + 3c_2\wDM\;,\\
    \wDM' &= \wDM\big(u^2 - c_2u - 3 + 3\wDM\big)\;.
\end{align}
\end{subequations}
The Jacobian of this system at the fixed point S has the trace and determinant quoted in \cref{eq:trdet_fp}.

Finally, we record how the system is modified by the production of dark matter from the SM brane considered in \cref{sec:production}. KK gravitons are produced in collisions of the thermal SM particles on the brane, so the energy which enters the tower is drawn from the SM radiation. Writing the production rate of \cref{eq:prod_rate} as $S = \kappa' T^8/M$ and using $\rho_{\m{r}} = (\pi^2/30)\,g_*(T)\,T^4$, with $g_*$ the number of relativistic degrees of freedom on the brane so that $T^4 = \rho_{\m{r}}/g$ with $g = \pi^2 g_*/30$ and $\rho_{\m{r}} = 3H^2\,\wR$, the fluid equations become
\begin{equation}
    \dot\rho_{\m{DM}} = -3H\rho_{\m{DM}} + \rho_{\m{DM}}\,\dot\phi\,\frac{\d\log M}{\d\phi} + S\;,\qquad \dot\rho_{\m{r}} = -4H\rho_{\m{r}} - S\;.
\end{equation}
The same steps as above give
\begin{subequations}\label{eq:aut_sys_prod}
\begin{align}
    u' &= -\frac{3}{2}\big(1 - \weff\big)u + 3\,c_1\,\wV + 3\,c_2\,\wDM\;,\\
    \wDM' &= \big(3\,\weff - c_2\,u\big)\wDM + s\;,\\
    \wB' &= 3\,\weff\,\wB\;,\\
    \wR' &= \big(3\,\weff - 1\big)\wR - s\;,\\
    \chi' &= \chi\Big(\frac{H'}{H} - \frac{M'}{M}\Big) = \chi\Big[\,c_2u - \frac{3}{2}\big(1+\weff\big)\Big]\;,
\end{align}
\end{subequations}
with
\begin{equation}
    s \equiv \frac{S}{3H^3} = 3\kappa\,\wR^2\,\chi\;,\qquad \chi \equiv \frac{H}{M}\;,\qquad \kappa = \frac{\kappa'}{g^2}\;.
\end{equation}
The equation for $u$ is unchanged, since the produced KK gravitons enter the radion equation only through $\rho_{\m{DM}}\,\d\log M/\d\phi$, which is already present. The source appears with opposite signs in the dark matter and radiation equations, which is what keeps the constraint \cref{eq:constraint} preserved by the flow. The variable $\chi$ is needed because $s$ depends on $H/M$ and not only on the abundances. The treatment of \cite{Gonzalo:2022jac} corresponds to the limit in which the sink in the radiation equation is negligible. The fixed points with $s \neq 0$ follow by the same product structure as before. The condition $\chi' = 0$ with $\chi \neq 0$ gives $c_2 u = \tfrac{3}{2}(1 + \weff)$, the condition $\wR' = 0$ gives $s = \wR(3\weff - 1)$, and $\wDM' = 0$ then gives $\wDM = 2s/[3(1 - \weff)]$, which is positive only for $\weff < 1$. These are the conditions \cref{eq:prod_pins}. The baryons must vanish since $\weff \neq 0$, the potential condition $c_1 u = 3(1+\weff)$ gives $c_1 = 2c_2$ if $\wV \neq 0$, and $u' = 0$ together with the constraint fixes the remaining variables. $\chi = (3\weff - 1)/(3\kappa\wR)$ is the value of $H/M$ at which production and dilution balance.

\section{Quasi-stasis linear response}\label{app:linresp}
On the branch $c_1 > 0$, the baryon direction is the slow unstable mode of S, with $\wB \propto e^{\lb N}$ and $\lb = 3\sweff$. We write the state within the face as
\begin{equation}
    (u,\, \wDM) = (u_s,\, \swDM) + \vec{k}\,\wB + \mathcal{O}(\wB^2)\;,
\end{equation}
and linearize the full system about S. The requirement that the correction follows the slow mode gives
\begin{equation}
    \big(J - \lb\,\mathbf{1}\big)\vec{k} = -\vec{S}\;,
\end{equation}
where $J$ is the Jacobian of the reduced system \cref{eq:aut_sys2}, whose trace and determinant are given in \cref{eq:trdet_fp}, and
\begin{equation}
    \vec{S} = \Big(\frac{\partial u'}{\partial \wB},\, \frac{\partial \wDM'}{\partial \wB}\Big)\bigg|_{\m{S}}
    = \Big(\frac{3}{2}u_s - 3c_1\;,\; 3\swDM\Big)\;.
\end{equation}
Solving, we find
\begin{equation}
    k_u = \frac{9c_2}{D}\;,\qquad
    k_\Omega = -\frac{(c_1^2 - c_1c_2 - 3)(2c_1^2c_2 - 2c_1c_2^2 - 3c_1 - 3c_2)}{(c_1-c_2)\,D}\;,
\end{equation}
with
\begin{equation}
    D = 2c_1^3c_2 - 4c_1^2c_2^2 - 3c_1^2 + 2c_1c_2^3 - 6c_1c_2 + 6c_2^2 + 9\;.
\end{equation}
The expansion is controlled by the instantaneous value of $\wB$ and by the ratio $\lb/\gamma$. It agrees with full integrations to better than a percent at $\wB \sim 10^{-3}$ and to a few percent at $\wB \sim 0.15$. The same formalism gives the adiabatic drift of \cref{sec:theorem} when the source vector is replaced by the term generated by the running of the slope.

\section{The thawing bound}\label{app:thawing}
In the anchored cosmology of \cref{sec:anchoring} the field $\phi$ is frozen deep in matter domination and thaws as the dark energy grows. We keep the terms linear in $u$ in \cref{eq:aut_u} and drop $u^2$ elsewhere, which means $\wDE \simeq \wV$ and $w_\phi \simeq -1$, we neglect radiation, and we write the dust as $\wDM = (1 - F_{\m{b}})(1 - \wDE)$ and $\wB = F_{\m{b}}(1 - \wDE)$, which ignores the change of the dark matter fraction by the mass drift. With $\weff \simeq -\wDE$ the friction coefficient is $\frac{3}{2}(1 + \wDE)$, and the sources are the slope of the potential and the coupled fraction of the dust,
\begin{equation}\label{eq:thaw_lin}
    \frac{\d u}{\d N} + \frac{3}{2}(1 + \wDE)\,u = 3c_1\,\wDE + 3c_2(1 - F_{\m{b}})(1 - \wDE)\;.
\end{equation}
We change variables using
\begin{equation}
    \frac{\d\wDE}{\d N} = 3\wDE(1-\wDE)
\end{equation}
which is the evolution of a cosmological constant in a dust background. This turns \cref{eq:thaw_lin} into a first order linear equation in $\wDE$,
\begin{equation}
    \frac{\d u}{\d\wDE} + \frac{1 + \wDE}{2\wDE(1 - \wDE)}\,u = \frac{c_1}{1 - \wDE} + \frac{c_2(1 - F_{\m{b}})}{\wDE}\;,
\end{equation}
whose integrating factor is $\sqrt{\wDE}/(1-\wDE)$. The general solution is a particular solution plus $C(1 - \wDE)/\sqrt{\wDE}$, the solution of the homogeneous equation in which the field coasts under Hubble friction alone. In matter domination this term is a velocity decaying as $a^{-3/2}$, the transient of the frozen field, so the anchored solution is the particular solution with $C = 0$ (no initial velocity for the frozen field $\phi$ relative to the attractor) which corresponds to the lower limit zero below. This is also why the result does not depend on the details of the anchoring at matter-radiation equality. This gives
\begin{align}
    u(\wDE) &= \frac{1-\wDE}{\sqrt{\wDE}}\int_0^{\wDE}
    \bigg[\frac{c_1}{1-x} + \frac{c_2(1-F_{\m{b}})}{x}\bigg]\frac{\sqrt{x}}{1-x}\,\d x\\
    &= c_1\,\big[1 - T(\wDE)\big] + 2c_2(1-F_{\m{b}})\,T(\wDE)\;,
\end{align}
with
\begin{equation}
    T(\wDE) = \frac{(1-\wDE)\,\m{arctanh}\sqrt{\wDE}}{\sqrt{\wDE}}\;,
\end{equation}
which is \cref{eq:thawing_u}. Since $T \rightarrow 1$ as $\wDE \rightarrow 0$ and $T \rightarrow 0$ as $\wDE \rightarrow 1$, the first term starts at zero and grows as the potential accelerates the field, and at $c_2 = 0$ it is the thawing expansion of \cite{Scherrer:2007pu}. The second term starts at $2c_2(1 - F_{\m{b}})$ and decays. During matter domination the coupled dust drives the field at the velocity $u = 2c_2$ of the point $\phi$MDE in \cref{tab:critpoints}, reduced by the coupled fraction of the dust, so a coupled radion is never exactly frozen but only slow. The approximation degrades once $u^2$ is no longer small compared to $\wDE$, which is why it fails at large $|c_1|$. Since dust is pressureless and radiation is negligible today,
\begin{equation}
    \weff(0) = \frac{u_0^2}{3} - \wDE^{(0)}\;,
\end{equation}
which gives the bound \cref{eq:floor}. Imposing $\weff(0) < -1/3$ gives the acceleration bound \cref{eq:accel_bound}. The formula is verified against full anchored integrations to better than a percent for $|c_1| \lesssim 1.2$.

\end{appendix}
\bibliographystyle{JHEP}
\bibliography{references}

\end{document}